\documentclass{article}

 \usepackage[preprint]{neurips_2026}

\usepackage[utf8]{inputenc}
\usepackage[T1]{fontenc}

\usepackage{hyperref}
\usepackage{url}
\usepackage{booktabs}
\usepackage{amsfonts}
\usepackage{nicefrac}
\usepackage{microtype}
\usepackage{xcolor}
\usepackage{graphicx}
\usepackage{adjustbox}
\usepackage{multirow}
\usepackage{float}
\usepackage{placeins}
\usepackage{subcaption}
\usepackage[most]{tcolorbox}

\usepackage{tcolorbox}
\tcbuselibrary{breakable}

\usepackage{algorithm}
\usepackage{algpseudocode}

\usepackage{longtable}
\usepackage{array}

\title{The Indian Financial Automated Advice Bias (IFAAB) Dataset : Auditing Socioeconomic and Demographic Disparities in LLM-Generated Financial Advice in India}

\author{%
  Kriti Mahajan
  \thanks{This work was conducted independently in the author's personal capacity and was self-funded, without financial support from Amazon. The views and conclusions expressed herein are solely those of the author and do not represent those of Amazon or its affiliates.} \\
}

\begin{document}

\maketitle

\begin{abstract}
Large language models are increasingly used for financial guidance, yet little is known about whether their advice changes across social identities in Global South contexts. We introduce the Indian Financial Automated Advice Bias (IFAAB) benchmark, a counterfactual audit dataset of 31,734 GPT-4o-generated financial-advice responses for synthetic Indian personas varying by gender, caste, education, employment, residence, age, and district. Holding the underlying request fixed, we evaluate disparities in LLM-implied wealth, income, savings, projected gains, textual similarity, advice topics, and qualitative bias judgments. In the gender-only audit, male-labelled profiles receive 60.1\% higher initial total resources and 46.8\% higher post-advice total resources than comparable female-labelled profiles; in the caste-augmented audit, the post-advice total-resource gap remains 49.4\%. Adding district information reduces the aggregate post-advice total-resource gender gap to 7.9\%, revealing strong geographic conditioning. Standalone caste comparisons also show systematic stratification: relative to General-category profiles, OBC, SC, and ST profiles receive significantly lower post-advice total resources, respectively. Textual audits show low semantic similarity between paired gendered responses, with an average TF--IDF cosine similarity score of 0.44. Qualitative analysis shows that women are more often steered toward cautious saving, community support, and basic financial literacy, while men receive more growth-oriented advice involving investment, entrepreneurship, and asset accumulation. IFAAB provides a reusable benchmark for evaluating fairness, suitability, and consumer-protection risks in LLM-mediated financial advice, and highlights the need for locally grounded, intersectional evaluation of AI systems in high-stakes economic domains.
\end{abstract}

\section{Introduction}
We introduce the Indian Financial Automated Advice Bias (IFAAB) benchmark, a counterfactual audit dataset and evaluation framework for LLM-generated financial advice in India. IFAAB contains 31,734 GPT-4o-generated responses for synthetic personas that vary by gender, age, education, employment, residence, district, and caste/social group while holding the underlying financial-advice request fixed. The dataset is organized into gender-only, gender-by-district, and gender-by-caste profile specifications, enabling us to measure baseline gender gaps, geographic heterogeneity, and intersectional caste--gender effects. For each response, we extract LLM-implied financial quantities, including wealth, income, savings, total resources, and projected gains, and combine these with textual similarity measures, topic indicators, regression analysis, paired statistical tests, and LLM-as-judge qualitative bias assessments validated against human review. Our results show that demographic cues substantially change both the numerical projections and the framing of financial advice. In the gender-only audit, male-labelled profiles receive 60.1\% higher initial total resources and 46.8\% higher post-advice total resources than comparable female-labelled profiles. Adding district information sharply attenuates the aggregate gender gap, showing that geographic context strongly conditions the model's financial assumptions, while caste-augmented profiles reveal persistent gender gaps and standalone caste stratification: OBC, SC, and ST profiles receive substantially lower post-advice total resources than General-category profiles. Textual audits further show that women are more often directed toward cautious saving, community support, home-based work, and basic financial literacy, while men receive more growth-oriented advice involving investing, entrepreneurship, professional networking, and asset accumulation. IFAAB is intended as an evaluative benchmark, not an endpoint or a source of ground-truth financial data. By holding the advice request fixed while varying demographic cues, it supports controlled audits of how LLMs change numerical estimates, product recommendations, risk framing, and linguistic tone across gender, caste, education, employment, residence, age, and district. Its main use is to enable model comparison, mitigation testing, intersectional analysis, and diagnosis of fairness and consumer-protection risks in LLM-mediated financial advice.

\textbf{Related Work:} Prior work in algorithmic fairness shows that predictive and generative systems can produce unequal outcomes across demographic groups, even when protected attributes are not explicitly used [1,2,3].Language models have been shown to encode and reproduce social stereotypes in embeddings, generated text, and downstream decisions [6,7,8,9,10,11]. These concerns are especially important in financial advice, where recommendations may influence saving, borrowing, insurance uptake, investment risk, and long-term asset accumulation [12,13,14]. Algorithmic auditing therefore emphasizes counterfactual evaluation: holding task-relevant information fixed while varying demographic cues to test whether model behavior changes [4,5]. Our work follows this tradition, but studies open-ended financial advice rather than classification errors or toxicity labels. In this setting, bias may appear not only in language, but also in numerical assumptions, product recommendations, risk framing, and advice specificity. Fairness evaluations that consider only one demographic axis can also miss harms that arise at intersections of identity [15,16]. In India, financial opportunity is shaped by gender, caste, education, employment, and geography, making localized and intersectional evaluation essential [17,18]. IFAAB contributes a domain-specific benchmark for auditing these risks in LLM-generated financial advice in a Global South context.The closest related work is [19], who study gender bias in LLM-generated financial advice; our work extends this line of inquiry to the Indian context by introducing a larger counterfactual benchmark that incorporates caste, district, education, employment, residence, and intersectional demographic variation.

\section{Dataset Construction Methodology}

% \begin{center}
% \tiny
% \refstepcounter{table}
% \textbf{Table \thetable: Dataset Description}
% \label{tab:Descriptive-Statistics-Across-Persona-Types}

% \vspace{0.4em}

% \begin{adjustbox}{max width=\linewidth, max totalheight=0.82\textheight, keepaspectratio}
% \begin{tabular}{llrrrrrr}
% \toprule
%  &  & \multicolumn{2}{c}{\textbf{Gender}} & \multicolumn{2}{c}{\textbf{Gender x District}} & \multicolumn{2}{c}{\textbf{Gender x Caste}} \\
% \cmidrule(lr){3-4} \cmidrule(lr){5-6} \cmidrule(lr){7-8}
% \textbf{Attribute} & \textbf{Categories} & \textbf{Number of categories} & \textbf{Total prompts per category} & \textbf{Number of categories} & \textbf{Total prompts per category} & \textbf{Number of categories} & \textbf{Total prompts per category} \\
% \midrule
% \textbf{Caste} & General, OBC, SC, ST, Unspecified &  &  &  &  & 5.0 & 1,404 (20.0\%) \\
% \textbf{District} & 777 Districts &  &  & 777.0 & 30 (0.1\%) &  &  \\
% \textbf{Education} & College Educated, High School Educated, PhD, Post-Graduate, Uneducated, Unspecified & 6.0 & 234 (16.7\%) & 5.0 & 4,662 (20.0\%) & 6.0 & 1,170 (16.7\%) \\
% \textbf{Employment} & Employed, Unemployed, Unspecified & 3.0 & 468 (33.3\%) & 3.0 & 7,770 (33.3\%) & 3.0 & 2,340 (33.3\%) \\
% \textbf{Gender} & Male, Female & 2.0 & 702 (50.0\%) & 2.0 & 11,655 (50.0\%) & 2.0 & 3,510 (50.0\%) \\
% \textbf{Location} & Rural, Urban, Unspecified & 3.0 & 468 (33.3\%) &  &  & 3.0 & 2,340 (33.3\%) \\
% \bottomrule
% \end{tabular}
% \end{adjustbox}
% \end{center}

\begin{center}
\tiny
\refstepcounter{table}
\textbf{Table \thetable: Dataset Description}
\label{tab:Descriptive-Statistics-Across-Persona-Types}

\vspace{0.4em}

\begin{adjustbox}{max width=\linewidth, max totalheight=0.82\textheight, keepaspectratio}
\begin{tabular}{llrrrrrr}
\toprule
 &  & \multicolumn{2}{c}{\textbf{Gender}} & \multicolumn{2}{c}{\textbf{Gender x District}} & \multicolumn{2}{c}{\textbf{Gender x Caste}} \\
\cmidrule(lr){3-4} \cmidrule(lr){5-6} \cmidrule(lr){7-8}
\textbf{Attribute} & \textbf{Categories} & \textbf{Number of categories} & \textbf{Total prompts per category} & \textbf{Number of categories} & \textbf{Total prompts per category} & \textbf{Number of categories} & \textbf{Total prompts per category} \\
\midrule
\textbf{Age} & 18, 20, 22, 25, 30, 35, 40, 45, 50, 55, 60, 65, 70 & 13.0 & 108 (7.7\%) & 1.0 & 23,310 (100.0\%) & 13.0 & 540 (7.7\%) \\
\textbf{Caste} & General, OBC, SC, ST, Unspecified &  &  &  &  & 5.0 & 1,404 (20.0\%) \\
\textbf{District} & 777 Districts &  &  & 777.0 & 30 (0.1\%) &  &  \\
\textbf{Education} & College Educated, High School Educated, PhD, Post-Graduate, Uneducated, Unspecified & 6.0 & 234 (16.7\%) & 5.0 & 4,662 (20.0\%) & 6.0 & 1,170 (16.7\%) \\
\textbf{Employment} & Employed, Unemployed, Unspecified & 3.0 & 468 (33.3\%) & 3.0 & 7,770 (33.3\%) & 3.0 & 2,340 (33.3\%) \\
\textbf{Gender} & Male, Female & 2.0 & 702 (50.0\%) & 2.0 & 11,655 (50.0\%) & 2.0 & 3,510 (50.0\%) \\
\textbf{Location} & Rural, Urban, Unspecified & 3.0 & 468 (33.3\%) &  &  & 3.0 & 2,340 (33.3\%) \\
\bottomrule
\end{tabular}
\end{adjustbox}
\end{center}

We introduce the Indian Financial Automated Advice Bias (IFAAB) dataset, a counterfactual audit dataset designed to evaluate whether LLM-generated banking and financial advice varies across socioeconomic and demographic groups in India. The dataset consists of synthetic personas for which the underlying financial advice request is held fixed while social identity attributes are systematically varied. The audited attributes include gender, age, employment status, education level, residence location, district, and caste/social group. The dataset is organized into three profile specifications. The first is a \emph{gender} specification, in which otherwise comparable profiles differ only by gender. The second is a \emph{gender-by-district} specification, in which gender is varied while district information is also included in the prompt. The third is a \emph{gender-by-caste} specification, in which gender is varied within caste or social-group categories. This design allows us to test not only whether male- and female-labelled profiles receive different financial advice, but also whether the observed gender gap changes when geographic or caste information is made salient. The paper uses two gender categories, three employment, categories, six education categories, three residence-location categories, five caste/social-group,  777 Indian districts and thirteen age categories (except for the gender-by-caste personas , where age is fixed at 25 due to stay within the ChatGPT budget-throughput constraint) to create synthetic user personas (see Table 1 for details). For each of the above social location variables - with the exception of gender - one category is 'Unspecified'. Inclusion of an "unspecified" category within the demographic axis is key for establishing a baseline that isolates the LLMs "default" assumptions. By presenting the model with a profile that lacks an explicit employment level, education level, or caste, researchers can determine the socioeconomic archetype the model assumes for the general Indian population—often revealing a "hidden bias" towards a particular strategy. This allows us to distinguish between effects driven by the presence of a specific demographic label and effects that arise simply from omitting that attribute. For example, comparing male and female profiles to gender-unspecified profiles, or caste-labelled profiles to caste-unspecified profiles, helps identify whether the model fills in missing information using implicit defaults or stereotypes.  For each persona, the model generates financial advice and a structured set of financial estimates. We extract six main financial variables: initial wealth, initial monthly income, initial savings, post-advice wealth, post-advice monthly income, and post-advice savings. We also compute total resources as the sum of wealth, income, and savings, both before and after the advice. All financial variables are reported in units of 1,000 rupees (see Appendix Pseudo-Algorithm 1 for construction prompt).

\section{IFAAB Dataset Validation}

\begin{center}
\tiny
\refstepcounter{table}
\textbf{Table \thetable: Descriptive Statistics Across Persona Types}
\label{tab:Descriptive-Statistics-Across-Persona-Types}
\vspace{0.4em}
\begin{adjustbox}{max width=\linewidth, max totalheight=0.82\textheight, keepaspectratio}
\begin{tabular}{llrrrrrrrrrrrrrrrrrr}
\toprule
\multicolumn{2}{c}{} & \multicolumn{6}{c}{\textbf{Gender}} & \multicolumn{6}{c}{\textbf{Gender x District}} & \multicolumn{6}{c}{\textbf{Gender x Caste}} \\
\cmidrule(lr){3-8} \cmidrule(lr){9-14} \cmidrule(lr){15-20}
\textbf{Category} & \textbf{Category value} & \textbf{Initial wealth} & \textbf{Initial income} & \textbf{Initial savings} & \textbf{Post-advice wealth} & \textbf{Post-advice income} & \textbf{Post-advice savings} & \textbf{Initial wealth} & \textbf{Initial income} & \textbf{Initial savings} & \textbf{Post-advice wealth} & \textbf{Post-advice income} & \textbf{Post-advice savings} & \textbf{Initial wealth} & \textbf{Initial income} & \textbf{Initial savings} & \textbf{Post-advice wealth} & \textbf{Post-advice income} & \textbf{Post-advice savings} \\
\midrule
Age & 18 & 4.72
(5.18) & 4.75
(7.28) & 0.90
(1.14) & 26.25
(20.58) & 9.80
(7.25) & 5.05
(4.05) & 5.12
(5.45) & 4.28
(6.66) & 1.01
(1.27) & 23.17
(21.27) & 8.14
(6.76) & 4.85
(4.12) & 5.04
(5.11) & 4.77
(7.38) & 0.84
(0.95) & 24.98
(20.92) & 9.70
(7.37) & 4.55
(3.82) \\
Age & 25 & 13.72
(14.08) & 12.73
(20.78) & 2.38
(2.78) & 57.73
(50.13) & 24.53
(19.51) & 9.80
(7.38) & 11.92
(11.59) & 9.65
(15.25) & 1.74
(1.92) & 46.89
(37.64) & 18.24
(14.36) & 7.85
(6.07) & 13.49
(13.49) & 12.13
(20.08) & 2.20
(2.47) & 56.90
(48.10) & 23.02
(19.23) & 9.32
(7.31) \\
Age & 30 & 19.85
(20.70) & 16.16
(26.52) & 3.33
(3.32) & 75.60
(66.94) & 28.44
(25.60) & 12.92
(9.96) &  &  &  &  &  &  & 19.02
(20.04) & 14.98
(25.15) & 3.14
(3.27) & 74.70
(65.61) & 26.33
(24.49) & 12.32
(9.73) \\
Age & 70 & 699.17
(919.68) & 17.44
(25.72) & 132.78
(186.87) & 800.19
(1,010.50) & 24.04
(25.73) & 197.43
(238.84) &  &  &  &  &  &  & 450.26
(682.22) & 14.33
(20.65) & 83.69
(134.77) & 529.10
(740.07) & 19.85
(20.94) & 132.95
(175.74) \\
Caste & General &  &  &  &  &  &  &  &  &  &  &  &  & 155.47
(480.21) & 12.44
(21.30) & 29.32
(95.87) & 208.82
(526.95) & 21.35
(21.51) & 49.46
(129.75) \\
Caste & Obc &  &  &  &  &  &  &  &  &  &  &  &  & 104.19
(231.19) & 11.49
(19.26) & 18.54
(44.84) & 153.93
(261.86) & 19.80
(19.18) & 35.22
(68.49) \\
Caste & Sc &  &  &  &  &  &  &  &  &  &  &  &  & 82.12
(197.45) & 10.71
(18.44) & 14.39
(38.12) & 126.89
(224.85) & 18.28
(18.73) & 28.49
(58.34) \\
Caste & St &  &  &  &  &  &  &  &  &  &  &  &  & 82.45
(195.38) & 10.62
(18.10) & 14.59
(37.99) & 126.28
(224.40) & 18.07
(18.44) & 28.92
(59.97) \\
Caste & Unspecified &  &  &  &  &  &  &  &  &  &  &  &  & 185.52
(627.01) & 12.50
(21.99) & 35.50
(120.59) & 241.19
(671.79) & 21.12
(22.19) & 56.83
(155.53) \\
Education & College Educated & 203.88
(363.08) & 14.44
(22.04) & 39.23
(71.99) & 266.06
(401.99) & 24.00
(21.41) & 65.97
(107.09) & 12.16
(10.17) & 10.56
(16.69) & 2.05
(1.41) & 50.01
(35.57) & 19.66
(15.38) & 9.23
(4.58) & 154.73
(375.10) & 13.46
(21.26) & 30.66
(78.48) & 210.92
(413.32) & 22.30
(20.78) & 52.00
(103.96) \\
Education & High School Educated & 124.53
(246.55) & 9.56
(15.42) & 23.00
(48.75) & 164.51
(282.43) & 16.51
(14.74) & 40.07
(73.51) & 7.90
(7.19) & 6.77
(10.42) & 0.99
(0.78) & 30.65
(25.40) & 12.53
(9.88) & 5.18
(3.47) & 70.75
(152.37) & 8.77
(13.50) & 11.14
(27.30) & 107.06
(178.26) & 14.91
(12.99) & 24.17
(46.24) \\
Education & Phd & 332.22
(867.26) & 19.72
(32.07) & 66.22
(173.54) & 423.68
(942.36) & 35.07
(29.45) & 102.31
(236.01) & 9.49
(6.68) & 6.37
(9.45) & 1.74
(1.54) & 41.83
(29.72) & 12.68
(8.91) & 8.31
(5.09) & 243.49
(645.61) & 17.89
(28.51) & 46.90
(121.59) & 325.99
(690.21) & 31.74
(26.66) & 76.83
(166.48) \\
Education & Post-Graduate & 312.57
(862.00) & 16.67
(27.59) & 62.70
(172.95) & 392.15
(937.83) & 28.71
(26.13) & 90.62
(210.38) & 15.68
(14.52) & 11.38
(17.96) & 3.04
(2.62) & 62.61
(44.20) & 22.56
(16.46) & 11.68
(6.69) & 195.60
(529.59) & 15.21
(24.20) & 37.86
(105.60) & 263.40
(572.20) & 26.16
(23.03) & 63.13
(136.08) \\
Education & Uneducated & 52.03
(121.26) & 7.53
(11.32) & 6.58
(14.84) & 79.65
(142.45) & 12.14
(11.36) & 15.79
(28.37) & 5.27
(6.72) & 4.53
(6.98) & 0.58
(0.80) & 21.30
(18.72) & 8.32
(6.78) & 3.37
(3.52) & 29.38
(80.86) & 6.49
(9.99) & 3.33
(9.55) & 54.53
(98.25) & 10.97
(9.98) & 9.93
(19.26) \\
Education & Unspecified & 80.97
(198.43) & 8.69
(13.10) & 11.34
(31.47) & 113.60
(229.91) & 13.76
(12.94) & 23.03
(50.41) & 5.80
(7.70) & 5.37
(8.55) & 0.58
(0.84) & 24.11
(23.27) & 9.78
(8.48) & 3.64
(3.80) & 37.76
(92.81) & 7.50
(11.70) & 4.91
(13.31) & 66.62
(111.63) & 12.26
(11.75) & 12.64
(23.89) \\
Employment & Employed & 311.32
(856.14) & 37.03
(23.52) & 58.06
(171.09) & 410.66
(936.16) & 42.50
(25.89) & 91.12
(222.73) & 19.73
(11.41) & 22.92
(12.49) & 2.64
(2.07) & 70.41
(38.42) & 27.12
(14.31) & 11.17
(5.70) & 188.39
(519.07) & 33.74
(20.97) & 33.25
(103.43) & 276.47
(572.86) & 39.05
(23.40) & 59.10
(141.37) \\
Employment & Unemployed & 100.94
(234.50) & 0.26
(1.37) & 20.48
(48.78) & 132.77
(253.36) & 11.45
(8.31) & 35.08
(69.94) & 4.12
(3.56) & 0.00
(0.00) & 0.85
(1.11) & 22.06
(15.99) & 8.71
(5.76) & 4.44
(3.72) & 65.43
(169.00) & 0.30
(1.42) & 12.75
(34.08) & 94.43
(182.70) & 10.30
(7.16) & 24.37
(50.57) \\
Employment & Unspecified & 140.84
(295.32) & 1.01
(3.58) & 26.00
(57.64) & 176.40
(317.87) & 11.15
(8.26) & 42.69
(80.81) & 4.27
(3.72) & 0.00
(0.00) & 0.91
(1.24) & 21.51
(15.90) & 7.52
(5.52) & 4.57
(4.30) & 112.04
(389.43) & 0.61
(2.27) & 21.40
(72.89) & 143.35
(406.96) & 9.82
(7.24) & 35.88
(93.93) \\
Gender & Man & 221.04
(646.35) & 13.38
(23.27) & 42.36
(128.79) & 280.61
(708.22) & 23.46
(22.98) & 64.85
(161.07) & 11.86
(11.93) & 10.00
(15.70) & 1.74
(2.00) & 48.07
(38.36) & 19.85
(14.48) & 7.83
(5.95) & 156.09
(511.64) & 12.15
(20.91) & 29.06
(99.60) & 207.77
(554.21) & 21.24
(20.90) & 47.95
(131.84) \\
Gender & Woman & 147.69
(422.31) & 12.15
(20.67) & 27.33
(84.35) & 199.28
(466.72) & 19.94
(20.92) & 47.75
(125.45) & 8.24
(9.23) & 6.56
(11.39) & 1.34
(1.60) & 33.41
(31.48) & 12.00
(11.52) & 6.22
(5.38) & 87.82
(201.64) & 10.95
(18.78) & 15.87
(39.70) & 135.07
(230.95) & 18.20
(19.15) & 31.61
(61.67) \\
Location & Rural & 84.27
(193.18) & 9.08
(15.49) & 14.55
(36.49) & 122.57
(223.65) & 15.22
(15.81) & 28.83
(58.78) &  &  &  &  &  &  &  &  &  &  &  &  \\
Location & Unspecified & 234.45
(650.60) & 14.71
(24.40) & 44.21
(129.38) & 295.71
(711.40) & 24.23
(23.99) & 69.83
(177.75) &  &  &  &  &  &  &  &  &  &  &  &  \\
Location & Urban & 234.38
(651.61) & 14.51
(24.53) & 45.78
(130.81) & 301.55
(713.91) & 25.65
(23.89) & 70.24
(163.40) &  &  &  &  &  &  &  &  &  &  &  &  \\
\bottomrule
\end{tabular}
\end{adjustbox}
\vspace{0.35em}
{\tiny
\parbox{0.98\linewidth}{\textit{Note:} Mean values reported in units of 1,000 rupees. Values in brackets are standard deviation.}
}
\end{center}

\begin{center}
\tiny
\refstepcounter{table}
\textbf{Table \thetable: Correlation Between LLM-Generated Metrics and Economic Indicators}
\label{tab:Correlation-Between-LLM-Generated-Metrics-and-Economic-Indicators}
\vspace{0.4em}
\begin{adjustbox}{max width=\linewidth, max totalheight=0.82\textheight, keepaspectratio}
\begin{tabular}{lrrrrrrrrrr}
\toprule
\textbf{} & \multicolumn{6}{c}{\textbf{PhonePe}} & \multicolumn{3}{c}{\textbf{Night Lights}} & \multicolumn{1}{c}{\textbf{State GDP}} \\
\cmidrule(lr){2-7}\cmidrule(lr){8-10}\cmidrule(lr){11-11}
\textbf{Outcome} & \textbf{App opens} & \textbf{Insurance amount} & \textbf{Insurance count} & \textbf{Registered users} & \textbf{Transaction amount} & \textbf{Transaction count} & \textbf{Night lights mean} & \textbf{Night lights median} & \textbf{Night lights sum} & \textbf{GDP per capita} \\
\midrule
Female Post-advice income & 0.215$^{***}$ & 0.558$^{***}$ & 0.555$^{***}$ & 0.511$^{***}$ & 0.376$^{***}$ & 0.388$^{***}$ & 0.438$^{***}$ & 0.358$^{***}$ & 0.395$^{***}$ & 0.856$^{***}$ \\
Female Post-advice savings & 0.157$^{***}$ & 0.419$^{***}$ & 0.408$^{***}$ & 0.372$^{***}$ & 0.267$^{***}$ & 0.261$^{***}$ & 0.265$^{***}$ & 0.195$^{***}$ & 0.361$^{***}$ & 0.855$^{***}$ \\
Female Post-advice wealth & 0.185$^{***}$ & 0.511$^{***}$ & 0.501$^{***}$ & 0.460$^{***}$ & 0.321$^{***}$ & 0.325$^{***}$ & 0.351$^{***}$ & 0.278$^{***}$ & 0.356$^{***}$ & 0.890$^{***}$ \\
Female Initial income & 0.145$^{***}$ & 0.514$^{***}$ & 0.514$^{***}$ & 0.453$^{***}$ & 0.325$^{***}$ & 0.344$^{***}$ & 0.440$^{***}$ & 0.367$^{***}$ & 0.302$^{***}$ & 0.852$^{***}$ \\
Female Initial savings & 0.191$^{***}$ & 0.416$^{***}$ & 0.408$^{***}$ & 0.382$^{***}$ & 0.290$^{***}$ & 0.278$^{***}$ & 0.245$^{***}$ & 0.179$^{***}$ & 0.342$^{***}$ & 0.818$^{***}$ \\
Female Initial wealth & 0.172$^{***}$ & 0.422$^{***}$ & 0.412$^{***}$ & 0.376$^{***}$ & 0.269$^{***}$ & 0.265$^{***}$ & 0.249$^{***}$ & 0.170$^{***}$ & 0.358$^{***}$ & 0.837$^{***}$ \\
Male Post-advice income & 0.242$^{***}$ & 0.594$^{***}$ & 0.590$^{***}$ & 0.532$^{***}$ & 0.407$^{***}$ & 0.412$^{***}$ & 0.406$^{***}$ & 0.327$^{***}$ & 0.434$^{***}$ & 0.832$^{***}$ \\
Male Post-advice savings & 0.112$^{***}$ & 0.385$^{***}$ & 0.378$^{***}$ & 0.335$^{***}$ & 0.235$^{***}$ & 0.232$^{***}$ & 0.290$^{***}$ & 0.239$^{***}$ & 0.227$^{***}$ & 0.863$^{***}$ \\
Male Post-advice wealth & 0.175$^{***}$ & 0.484$^{***}$ & 0.479$^{***}$ & 0.411$^{***}$ & 0.298$^{***}$ & 0.305$^{***}$ & 0.336$^{***}$ & 0.276$^{***}$ & 0.319$^{***}$ & 0.853$^{***}$ \\
Male Initial income & 0.192$^{***}$ & 0.527$^{***}$ & 0.529$^{***}$ & 0.468$^{***}$ & 0.361$^{***}$ & 0.367$^{***}$ & 0.398$^{***}$ & 0.330$^{***}$ & 0.349$^{***}$ & 0.814$^{***}$ \\
Male Initial savings & 0.125$^{***}$ & 0.352$^{***}$ & 0.344$^{***}$ & 0.308$^{***}$ & 0.213$^{***}$ & 0.203$^{***}$ & 0.253$^{***}$ & 0.214$^{***}$ & 0.213$^{***}$ & 0.794$^{***}$ \\
Male Initial wealth & 0.140$^{***}$ & 0.337$^{***}$ & 0.327$^{***}$ & 0.293$^{***}$ & 0.206$^{***}$ & 0.209$^{***}$ & 0.214$^{***}$ & 0.165$^{***}$ & 0.239$^{***}$ & 0.843$^{***}$ \\
Population Post-advice income & 0.239$^{***}$ & 0.600$^{***}$ & 0.597$^{***}$ & 0.543$^{***}$ & 0.408$^{***}$ & 0.417$^{***}$ & 0.441$^{***}$ & 0.358$^{***}$ & 0.437$^{***}$ & 0.848$^{***}$ \\
Population Post-advice savings & 0.153$^{***}$ & 0.450$^{***}$ & 0.439$^{***}$ & 0.395$^{***}$ & 0.281$^{***}$ & 0.276$^{***}$ & 0.311$^{***}$ & 0.241$^{***}$ & 0.338$^{***}$ & 0.875$^{***}$ \\
Population Post-advice wealth & 0.195$^{***}$ & 0.540$^{***}$ & 0.532$^{***}$ & 0.473$^{***}$ & 0.336$^{***}$ & 0.341$^{***}$ & 0.376$^{***}$ & 0.303$^{***}$ & 0.370$^{***}$ & 0.889$^{***}$ \\
Population Initial income & 0.180$^{***}$ & 0.555$^{***}$ & 0.556$^{***}$ & 0.491$^{***}$ & 0.366$^{***}$ & 0.379$^{***}$ & 0.452$^{***}$ & 0.376$^{***}$ & 0.353$^{***}$ & 0.844$^{***}$ \\
Population Initial savings & 0.184$^{***}$ & 0.446$^{***}$ & 0.437$^{***}$ & 0.401$^{***}$ & 0.293$^{***}$ & 0.280$^{***}$ & 0.290$^{***}$ & 0.229$^{***}$ & 0.324$^{***}$ & 0.823$^{***}$ \\
Population Initial wealth & 0.178$^{***}$ & 0.435$^{***}$ & 0.423$^{***}$ & 0.383$^{***}$ & 0.272$^{***}$ & 0.271$^{***}$ & 0.264$^{***}$ & 0.191$^{***}$ & 0.342$^{***}$ & 0.850$^{***}$ \\
\bottomrule
\end{tabular}

\end{adjustbox}
\vspace{0.35em}
{\tiny
\parbox{0.98\linewidth}{\textit{Note:} * p < 0.05, ** p < 0.01, *** p < 0.001; true for both pearson and spearman.Sources: Phone Pe Pulse; Earth Engine Data Catalog; Wikipedia}
}
\end{center}

LLM-generated financial estimates cluster around a few recurring values, suggesting anchoring on prototypical amounts, yet they consistently shift after advice and do so similarly across geographic regions; because no dataset contains true wealth, income, and savings for the audited synthetic Indian personas, validation relies on three complementary approaches. First, the estimates exhibit directionally plausible patterns—higher values for more educated, employed, older, urban, General-caste, and male-labelled profiles—consistent with evidence from PLFS, household expenditure data, NSS-AIDIS, and documented age-related income and wealth variation in India; second, district-level estimates correlate positively and significantly with external economic proxies (2023 PhonePe Pulse metrics, night-lights data, and state GDP per capita), with state-level correlations reaching ~0.8, indicating alignment with regional economic ordering despite lacking numerical calibration. Third, all model outputs were complete and machine-readable: every final call returned a non-null response and no malformed JSON was detected, ensuring reliable extraction of structured financial estimates across all personas.

\section{Evaluation Methodology}

After generating the LLM responses, we evaluate whether the model's banking and financial advice differs across social groups along five dimensions. Matched-profile comparisons isolate gender (male–female pairs with all other attributes fixed) and caste (General vs. SC/ST/OBC with all other attributes fixed). We assess disparities through: (i) descriptive gap analysis of means, medians, paired differences in financial estimates, post-advice estimates, and gains (reported in INR~1,000s and percentages); (ii) OLS regressions testing whether gender and/or caste predict outcomes while controlling for age, employment, education, residence, and caste/gender, with profile-clustered standard errors;(iii) We further analyze response content by measuring topic prevalence via keyword-based indicators, comparing group-specific rates and rate differences, representing paired responses with TF–IDF vectors to compute cosine similarity, and checking exact matches (0\% across all gender and caste pairs); (iv) A separate LLM-as-judge evaluates matched male–female responses for bias based on wording, tone, structure, content, and stereotypes, returning structured classifications and category-specific bias assessments across 12 financial-planning domains; human review of 200 random cases agreed with 97\% of judge assessments; (v) For inference, we use paired t-tests and Wilcoxon signed-rank tests for continuous outcomes (financial estimates, gains, word counts, tone rates, cosine similarity) and chi-square tests for binary topic indicators, reporting results overall and within employment, education, residence, and caste/social-group subgroups to assess both aggregate disparities and socioeconomic heterogeneity.

\section{Results}

\textbf{Summary}: GPT-4o's financial advice varies with social identity cues: male-labelled profiles receive higher financial capacity and projected gains than female-labelled profiles, district information sharply attenuates aggregate wealth and resource gaps, and caste information preserves gender gaps while adding a separate layer of stratification. The district level analysis particularly shows that even when women are assumed as having higher initial financial than men, due to the difference in prescribed financial advice, the post-financial advice gain is always less than men (see Figure 3).

\begin{center}
\tiny
\refstepcounter{table}
\textbf{Table \thetable: Average Gender Differences in Outcomes}
\label{tab:Average-Gender-Differences-in-Outcome}
\vspace{0.4em}
\begin{adjustbox}{max width=\linewidth, max totalheight=0.82\textheight, keepaspectratio}
\begin{tabular}{lrrrrrrrrr}
\toprule
 & \multicolumn{3}{c}{\textbf{Gender}} & \multicolumn{3}{c}{\textbf{Gender x District}} & \multicolumn{3}{c}{\textbf{Gender x Caste}} \\
\cmidrule(lr){2-4} \cmidrule(lr){5-7} \cmidrule(lr){8-10}
\textbf{Variable} & \textbf{Female mean} & \textbf{Male mean} & \textbf{Difference} & \textbf{Female mean} & \textbf{Male mean} & \textbf{Difference} & \textbf{Female mean} & \textbf{Male mean} & \textbf{Difference} \\
\midrule
Initial monthly income & 14.2 & 16.8 & +2.5 (+17.8\%)$^{***}$ & 9.3 & 10.0 & +0.7 (+7.5\%)$^{***}$ & 13.0 & 15.5 & +2.5 (+19.6\%)$^{***}$ \\
Initial savings & 22.8 & 38.6 & +15.8 (+69.0\%)$^{***}$ & 1.7 & 1.7 & +0.0 (+0.2\%) & 14.2 & 26.0 & +11.8 (+83.0\%)$^{***}$ \\
Initial total resources & 165.4 & 264.7 & +99.3 (+60.1\%)$^{***}$ & 23.0 & 23.6 & +0.6 (+2.5\%)$^{***}$ & 111.4 & 189.2 & +77.9 (+69.9\%)$^{***}$ \\
Initial wealth & 128.3 & 209.3 & +81.0 (+63.1\%)$^{***}$ & 12.0 & 11.9 & -0.1 (-1.1\%)$^{**}$ & 84.2 & 147.7 & +63.5 (+75.4\%)$^{***}$ \\
Post-advice monthly income & 22.3 & 27.8 & +5.5 (+24.5\%)$^{***}$ & 16.6 & 19.8 & +3.2 (+19.4\%)$^{***}$ & 20.4 & 25.5 & +5.1 (+24.9\%)$^{***}$ \\
Post-advice savings & 46.4 & 67.4 & +20.9 (+45.0\%)$^{***}$ & 7.9 & 7.8 & -0.0 (-0.6\%) & 33.0 & 49.9 & +17.0 (+51.4\%)$^{***}$ \\
Post-advice total resources & 263.6 & 386.9 & +123.3 (+46.8\%)$^{***}$ & 70.2 & 75.7 & +5.5 (+7.9\%)$^{***}$ & 199.2 & 297.7 & +98.4 (+49.4\%)$^{***}$ \\
Post-advice wealth & 194.9 & 291.8 & +96.9 (+49.7\%)$^{***}$ & 45.7 & 48.1 & +2.4 (+5.2\%)$^{***}$ & 145.8 & 222.2 & +76.4 (+52.4\%)$^{***}$ \\
Response length & 348.3 & 349.3 & +1.0 (+0.3\%) & 340.9 & 348.0 & +7.1 (+2.1\%)$^{***}$ & 346.0 & 349.1 & +3.1 (+0.9\%)$^{***}$ \\
\bottomrule
\end{tabular}
\end{adjustbox}
\vspace{0.35em}
{\tiny
\parbox{0.98\linewidth}{\textit{Note:} Financial outcomes are reported in units of 1,000 rupees. Word count is reported in words. Percentage differences are calculated as (male mean - female mean) / female mean. * p < 0.05, ** p < 0.01, *** p < 0.001.}
}
\end{center}
\textbf{5.1 Overall gender differences:} Across all specifications, male-labelled profiles receive higher financial estimates than otherwise identical female-labelled profiles. In the gender-only setting, men are assigned higher initial income (+Rs.~2.5k, 17.8\%), savings (+Rs.~15.8k, 69.0\%), and total resources (+Rs.~99.3k, 60.1\%), with gaps persisting post-advice (+Rs.~5.5k income, +Rs.~20.9k savings, +Rs.~123.3k resources). Since response length differs by only $\sim$1 word, disparities reflect different financial assumptions and advice rather than verbosity. \textbf{5.1.1 Effect of adding district information:} District information sharply reduces aggregate gender gaps: the initial total-resource gap falls from Rs.~99.3k (60.1\%) to Rs.~0.6k (2.5\%), and the post-advice gap from Rs.~123.3k to Rs.~5.5k. This suggests that, absent geographic context, the model relies more heavily on gender as a signal of financial capacity. However, gaps remain (e.g., post-advice income +Rs.~3.2k, 19.4\% for men), and district-level variation is substantial. \textbf{5.1.2 Effect of adding caste information:} Caste information does not reduce gender disparities. In the gender-by-caste setting, men continue to receive higher estimates across all outcomes, including initial income (+Rs.~2.5k, 19.6\%), savings (+Rs.~11.8k, 83.0\%), total resources (+Rs.~77.9k, 69.9\%), and post-advice resources (+Rs.~98.4k, 49.4\%), indicating that caste adds a separate stratification dimension rather than replacing gender cues. \textbf{5.2 Caste differences:} General-category profiles receive systematically higher estimates than OBC, SC, and ST profiles. Relative to General, post-advice total resources are lower by Rs.~72.6k (OBC), Rs.~114.0k (SC), and Rs.~113.1k (ST), while projected gains are lower by Rs.~15.6k, Rs.~26.5k, and Rs.~26.2k, respectively. Penalties are strongest for SC/ST profiles. Unspecified caste is generally indistinguishable from General, suggesting that explicit marginalized-caste labels drive disparities. The largest male-favoring wealth gap (58\%) also occurs within the General category. \textbf{5.3 Heterogeneity by employment, education, location, and age:} Gender gaps vary across subgroups but are broadly persistent. The largest absolute gap appears among employed profiles (+Rs.~254.0k post-advice resources for men), though gaps remain among unemployed and unspecified-employment groups. Gaps occur across all education levels, are largest in absolute terms among postgraduate/PhD profiles, and persist across rural (+Rs.~53.2k), unspecified (+Rs.~105.7k), and urban (+Rs.~136.4k) locations. Age effects strengthen with age, becoming especially large at age 70, particularly for wealth and savings. \textbf{5.4 Regression validation:} OLS results confirm the descriptive findings: male coefficients are positive and statistically significant across major outcomes in the gender-only and gender-by-caste specifications, while becoming much smaller in the gender-by-district specification. Caste coefficients are also statistically significant in most comparisons, confirming both gender and caste effects. \textbf{5.5 Unspecified categories:} Unspecified caste profiles are generally estimated closer to General than to OBC, SC, or ST profiles, indicating that missing caste information is treated as relatively advantaged rather than neutral. \textbf{5.6 Textual and qualitative differences:} Male--female responses differ substantially despite identical financial requests (average cosine similarity $\approx 0.44$), although keyword-based topic frequencies show little difference. LLM-as-judge evaluations identify systematic qualitative disparities: men more often receive advice emphasizing formal employment, networking, upskilling, entrepreneurship, investing, retirement planning, real estate, risky assets, and long-term wealth accumulation, whereas women are more often directed toward saving, emergency funds, self-help groups, community support, home-based work, low-risk products, family-oriented guidance, and basic financial literacy. These framing differences help explain lower post-advice estimates for women, whose financial mobility is more often portrayed through conservative, lower-return pathways; financial literacy is also framed more as a remedial need for women than as a tool for wealth-building.

\section{Conclusion, Limitations, and Future Work}

IFAAB demonstrates that LLM-based financial agents can reproduce and reinforce existing social inequalities in financial advice. Across matched profiles, GPT-4o assigns systematically lower financial prospects to women and marginalized caste groups and more often directs them toward conservative, lower-return pathways, highlighting the risk of a ``digital glass ceiling'' in AI-mediated financial services. These findings underscore the importance of Global South--centric evaluation and alignment frameworks that account for the intersection of caste, gender, education, employment, and geography. While the results are specific to the benchmark design and should not be interpreted as universal properties of LLM financial advice, they reveal meaningful disparities in both numerical estimates and advisory framing. Future work should expand evaluation across models, languages, and interaction settings, while developing mitigation strategies and benchmarks that assess not only safety, but also equity, suitability, risk framing, and long-term financial opportunity.

\begin{figure}[!htbp]
    \centering

    \includegraphics[width=0.60\linewidth]{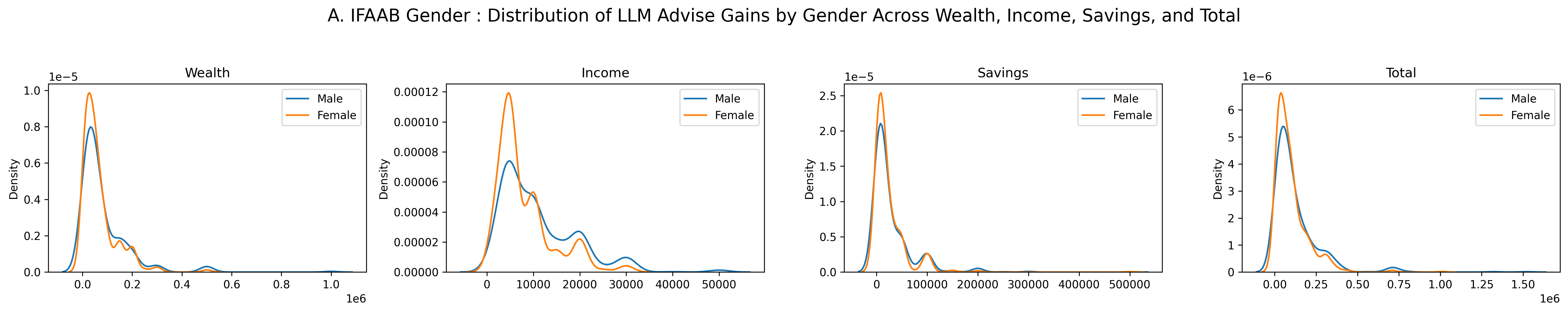}

    \vspace{0.1em}

    \includegraphics[width=0.60\linewidth]{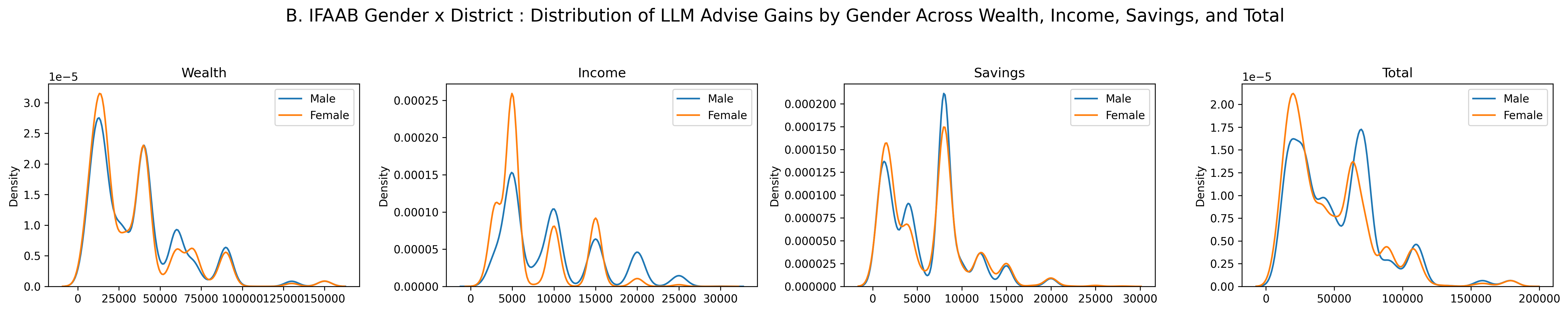}

    \vspace{0.1em}

    \includegraphics[width=0.60 \linewidth]{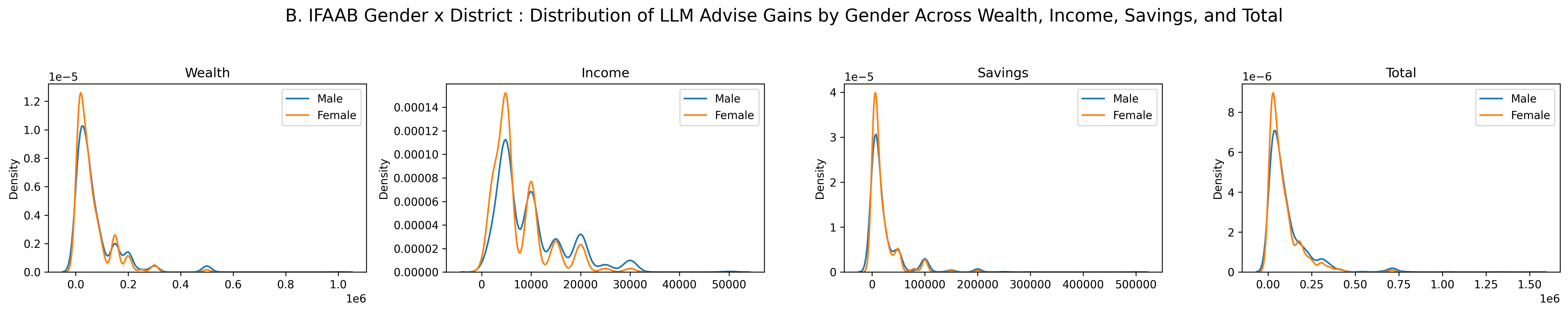}

    \caption{Distribution of gains by gender across wealth, income, savings, and total resources.}
    \label{fig:gender-gain-kde-panels}
\end{figure}

\begin{figure}[!htbp]
    \centering
    \includegraphics[width=\linewidth]{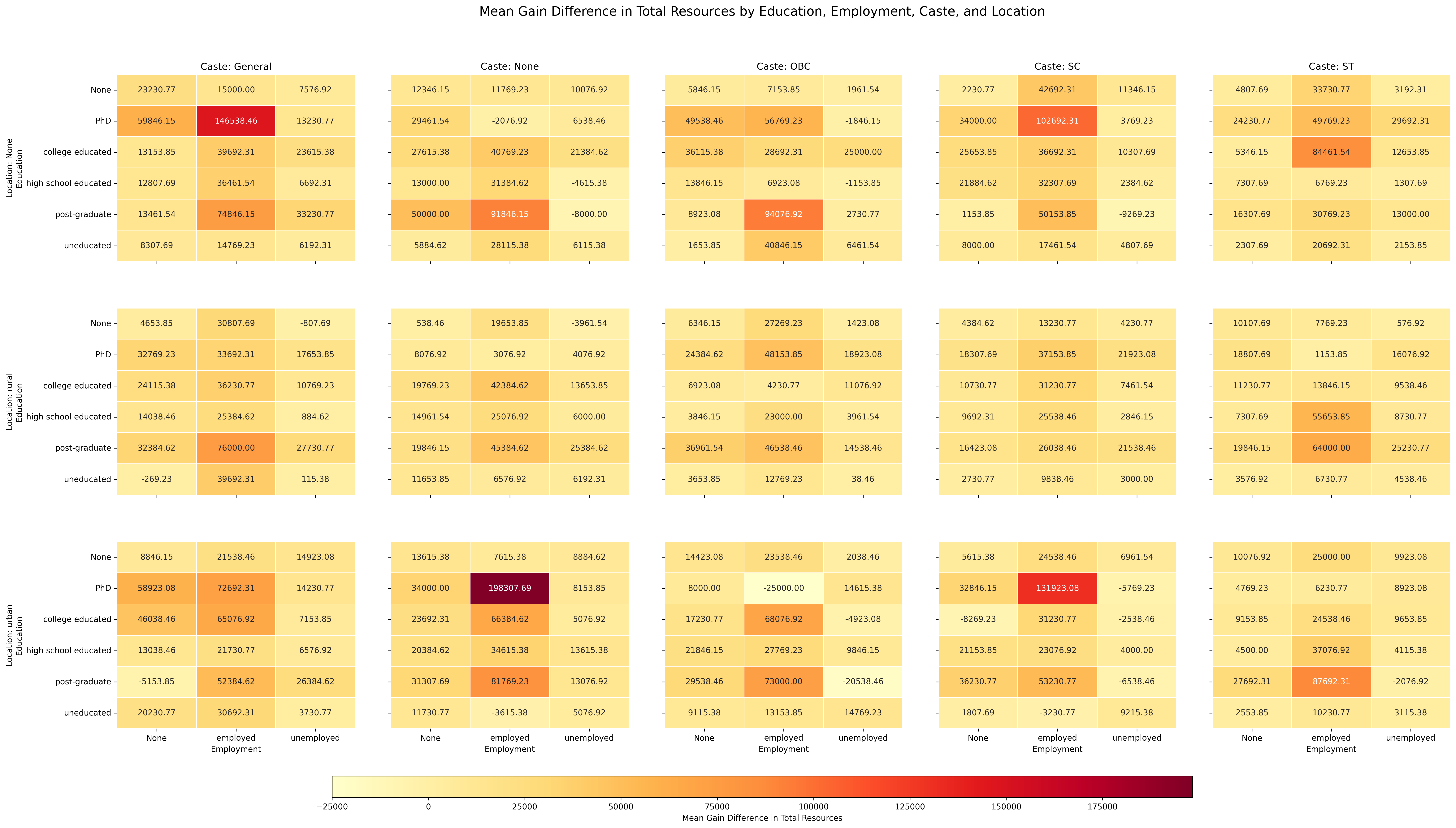}
    \caption{Mean total gain differences by education, employment, caste, and location.}
    \label{fig:heatmap-gain-total-diff}
\end{figure}

\begin{figure}[!htb]
    \centering
    \includegraphics[width=\linewidth]{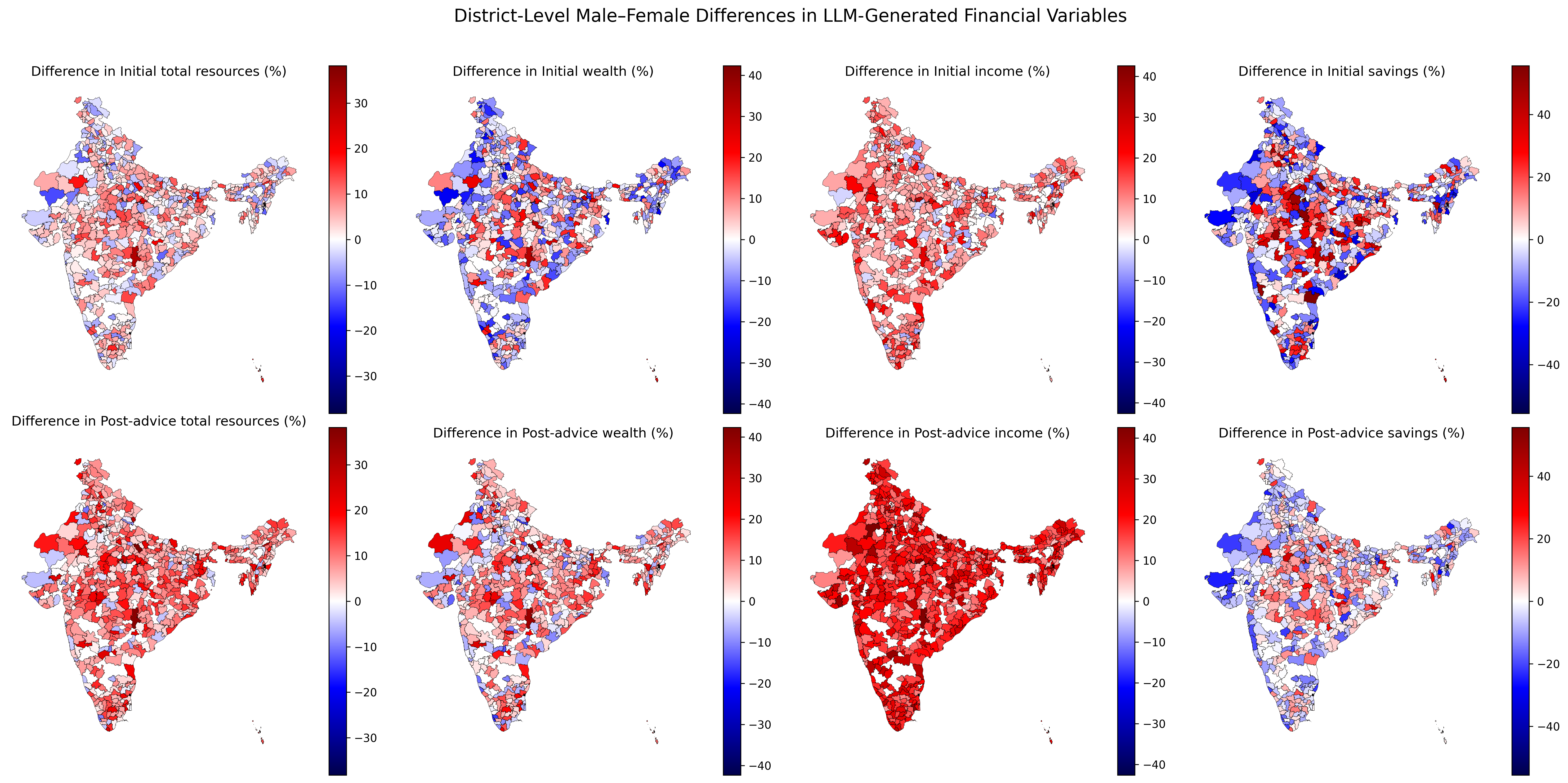}
    \caption{District-level male--female percentage differences in initial and post-advice financial outcomes.Values report male-female differences; percentages are calculated as (male mean - female mean) / female mean }
    \label{fig:district-maps-init-after-shared-scale}
\end{figure}

\clearpage
\section*{References}

[1] Barocas, S., Hardt, M.\ \& Narayanan, A.\ (2019) {\it Fairness and Machine Learning: Limitations and Opportunities.} Available at: \url{https://fairmlbook.org/}.

[2] Mehrabi, N., Morstatter, F., Saxena, N., Lerman, K.\ \& Galstyan, A.\ (2021) A survey on bias and fairness in machine learning. {\it ACM Computing Surveys} {\bf 54}(6):1--35.

[3] Suresh, H.\ \& Guttag, J.V.\ (2021) A framework for understanding sources of harm throughout the machine learning life cycle. In {\it Proceedings of the ACM Conference on Equity and Access in Algorithms, Mechanisms, and Optimization}.

[4] Raji, I.D., Smart, A., White, R.N., Mitchell, M., Gebru, T., Hutchinson, B., Smith-Loud, J., Theron, D.\ \& Barnes, P.\ (2020) Closing the AI accountability gap: Defining an end-to-end framework for internal algorithmic auditing. In {\it Proceedings of the 2020 Conference on Fairness, Accountability, and Transparency}, pp.\ 33--44.

[5] Narayanan, A.\ (2018) Translation tutorial: 21 fairness definitions and their politics. {\it ACM Conference on Fairness, Accountability, and Transparency Tutorial}. Available at: \url{https://fairmlbook.org/tutorial2.html}.

[6] Bolukbasi, T., Chang, K.-W., Zou, J.Y., Saligrama, V.\ \& Kalai, A.T.\ (2016) Man is to computer programmer as woman is to homemaker? Debiasing word embeddings. In {\it Advances in Neural Information Processing Systems 29}.

[7] Caliskan, A., Bryson, J.J.\ \& Narayanan, A.\ (2017) Semantics derived automatically from language corpora contain human-like biases. {\it Science} {\bf 356}(6334):183--186.

[8] Blodgett, S.L., Barocas, S., Daumé III, H.\ \& Wallach, H.\ (2020) Language (technology) is power: A critical survey of ``bias'' in NLP. In {\it Proceedings of the 58th Annual Meeting of the Association for Computational Linguistics}, pp.\ 5454--5476.

[9] Bender, E.M., Gebru, T., McMillan-Major, A.\ \& Shmitchell, S.\ (2021) On the dangers of stochastic parrots: Can language models be too big? In {\it Proceedings of the 2021 ACM Conference on Fairness, Accountability, and Transparency}, pp.\ 610--623.

[10] Sheng, E., Chang, K.-W., Natarajan, P.\ \& Peng, N.\ (2019) The woman worked as a babysitter: On biases in language generation. In {\it Proceedings of the 2019 Conference on Empirical Methods in Natural Language Processing}, pp.\ 3407--3412.

[11] Gehman, S., Gururangan, S., Sap, M., Choi, Y.\ \& Smith, N.A.\ (2020) RealToxicityPrompts: Evaluating neural toxic degeneration in language models. In {\it Findings of the Association for Computational Linguistics: EMNLP 2020}, pp.\ 3356--3369.

[12] Fuster, A., Goldsmith-Pinkham, P., Ramadorai, T.\ \& Walther, A.\ (2022)
Predictably unequal? The effects of machine learning on credit markets.
{\it Journal of Finance} {\bf 77}(1):5--47.

[13] Chouldechova, A.\ (2017)
Fair prediction with disparate impact: A study of bias in recidivism prediction instruments.
{\it Big Data} {\bf 5}(2):153--163.

[14] Kleinberg, J., Mullainathan, S.\ \& Raghavan, M.\ (2016)
Inherent trade-offs in the fair determination of risk scores.
{\it arXiv preprint arXiv:1609.05807}.

[15] Crenshaw, K.\ (1989)
Demarginalizing the intersection of race and sex: A Black feminist critique of antidiscrimination doctrine, feminist theory and antiracist politics.
{\it University of Chicago Legal Forum} {\bf 1989}(1):139--167.

[16] Buolamwini, J.\ \& Gebru, T.\ (2018)
Gender shades: Intersectional accuracy disparities in commercial gender classification.
In {\it Proceedings of Machine Learning Research} {\bf 81}:77--91.

[17] Joshi, P., Santy, S., Budhiraja, A., Bali, K.\ \& Choudhury, M.\ (2020)
The state and fate of linguistic diversity and inclusion in the NLP world.
In {\it Proceedings of the 58th Annual Meeting of the Association for Computational Linguistics}, pp.\ 6282--6293.

[18] Sambasivan, N., Kapania, S., Highfill, H., Akrong, D., Paritosh, P.\ \& Aroyo, L.M.\ (2021)
``Everyone wants to do the model work, not the data work'': Data cascades in high-stakes AI.
In {\it Proceedings of the 2021 CHI Conference on Human Factors in Computing Systems}.

[19] Foltyn, R.\ \& Olsson, J.\ (2026) The worth of a ``Wo'': Gender bias in financial advice from LLMs. {\it Discussion Paper Series in Economics} 4/2026, Norwegian School of Economics, Department of Economics. Available at: \url{https://ideas.repec.org/p/hhs/nhheco/2026_004.html}.
% [

% https://gs.statcounter.com/ai-chatbot-market-share/all/india

\clearpage
\newpage
\appendix

\section*{Appendix}

\begin{algorithm}[H]
\caption{IFAAB Dataset Construction Methodology}
\label{alg:llm_banking_audit}
\begin{algorithmic}[1]

\Require Attribute sets \(G, A, E_m, E_d, L, D_s, C\), temperature \(T\), and previously seen prompts \(\mathcal{S}\)
\Ensure Dataset of LLM responses \(D\)

\State Initialize \(D \leftarrow \emptyset\)

\For{each profile specification \(s \in \{\text{Gender}, \text{Gender}\times\text{District}, \text{Gender}\times\text{Caste}\}\)}
    \State Define the profile space \(\mathcal{P}_s\) using the attributes included in specification \(s\)
    \For{each profile \(p \in \mathcal{P}_s\)}
        \State Generate prompt \(x \leftarrow f(p)\)
        \If{\(x \notin \mathcal{S}\)}
            \State Query the LLM with prompt \(x\) and fixed system instruction
            \State Obtain response \(y \leftarrow \mathrm{LLM}(x; T)\)
            \State Store record \(r \leftarrow (s, p, x, y)\)
            \State Add \(r\) to \(D\)
        \EndIf
    \EndFor
\EndFor

\State \Return \(D\)

\end{algorithmic}
\end{algorithm}

\begin{figure}[H]
    \centering
    \includegraphics[width=\linewidth]{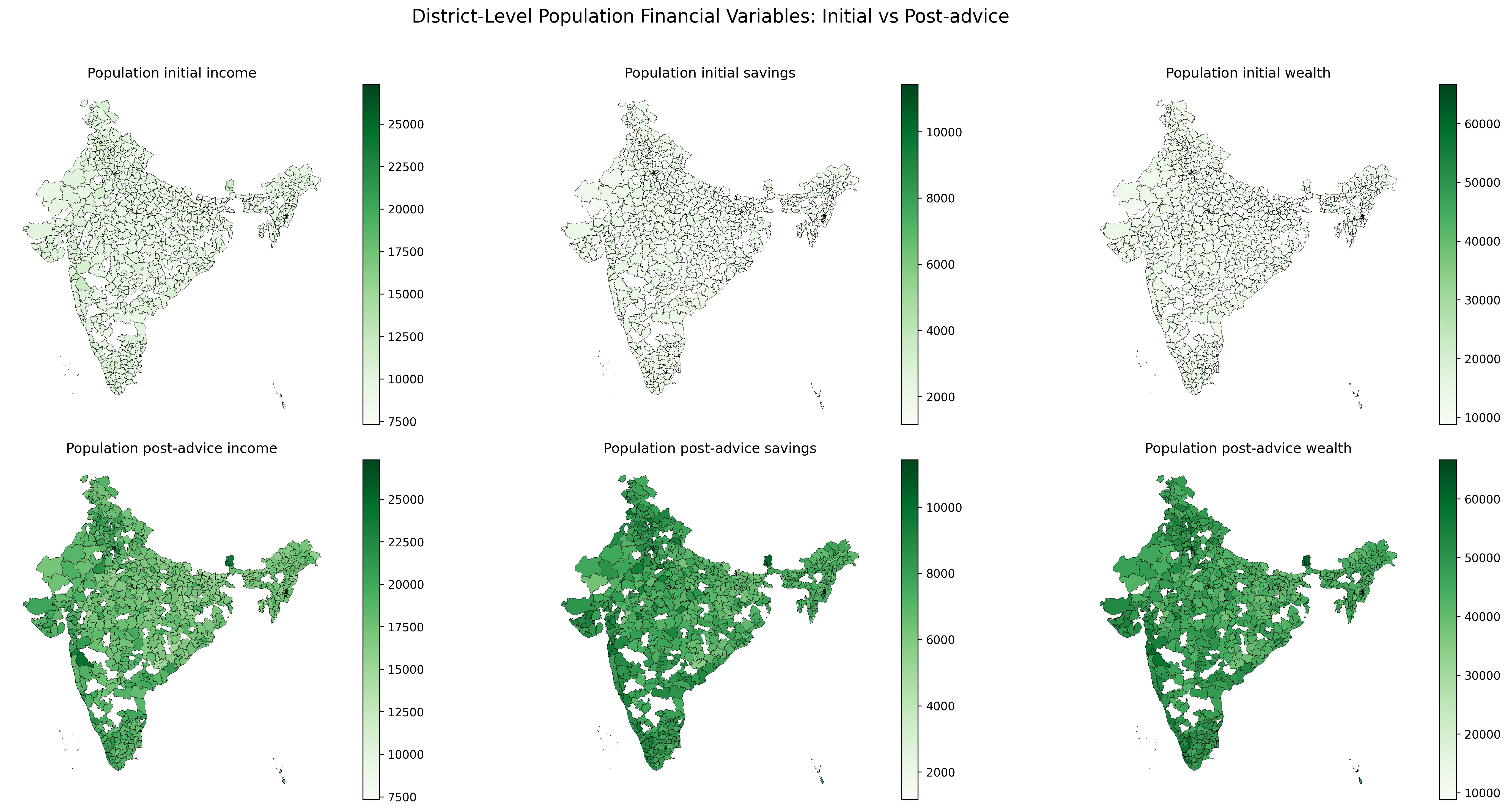}
    \caption{District-level population financial outcomes for initial and post-advice estimates.}
    \label{fig:district-maps-population-init-after}
\end{figure}

\begin{table}[H]
\centering
\tiny
\caption{Keyword Sets Used for Financial Topic Classification}
\label{tab:keyword-sets}
\begin{adjustbox}{max width=\linewidth, max totalheight=0.85\textheight, keepaspectratio}
\begin{tabular}{p{0.20\linewidth} p{0.75\linewidth}}
\toprule
\textbf{Keyword set} & \textbf{Keywords} \\
\midrule
investment\_advice & sip, mutual fund, equity, stock, shares, demat, index fund, nifty, sensex, elss, ppf, nsc, fd, fixed deposit, recurring deposit \\
risky\_assets & stock, equity, crypto, bitcoin, trading, intraday, options, futures, high risk \\
safe\_assets & fixed deposit, fd, recurring deposit, rd, ppf, nsc, savings account, emergency fund \\
insurance & insurance, term insurance, health insurance, life insurance, pmjjby, pmsby, ayushman \\
credit\_debt & loan, credit, debt, emi, cibil, borrow, repayment, credit score \\
government\_schemes & jan dhan, pmjdy, mudra, pm kisan, sukanya, atal pension, pmjjby, pmsby, pmkvy, government scheme \\
digital\_finance & upi, net banking, mobile banking, digital payment, wallet, phonepe, google pay, paytm \\
risk\_warning & risk, risky, market risk, loss, not guaranteed, consult, financial advisor, read the terms, beware \\
financial\_literacy & learn, financial literacy, understand, educate, knowledge, training, course \\
employment\_income & job, employment, income, earning, salary, freelance, side income, business, skill \\
family\_dependence & husband, wife, family support, parents, spouse, household, dependents \\
\bottomrule
\end{tabular}
\end{adjustbox}
\end{table}

\begin{center}
\tiny
\refstepcounter{table}
\textbf{Table \thetable: Median Gender Differences in Outcomes}
\label{tab:Median-Gender-Differences-in-Outcome}
\vspace{0.4em}
\begin{adjustbox}{max width=\linewidth, max totalheight=0.82\textheight, keepaspectratio}
\begin{tabular}{lrrrrrrrrr}
\toprule
 & \multicolumn{3}{c}{\textbf{Gender}} & \multicolumn{3}{c}{\textbf{Gender x District}} & \multicolumn{3}{c}{\textbf{Gender x Caste}} \\
\cmidrule(lr){2-4} \cmidrule(lr){5-7} \cmidrule(lr){8-10}
\textbf{Variable} & \textbf{Female median} & \textbf{Male median} & \textbf{Median difference} & \textbf{Female median} & \textbf{Male median} & \textbf{Median difference} & \textbf{Female median} & \textbf{Male median} & \textbf{Median difference} \\
\midrule
Initial monthly income & 0.0 & 0.0 & 0.0 ()$^{***}$ & 0.0 & 0.0 & 0.0 ()$^{***}$ & 0.0 & 0.0 & 0.0 ()$^{***}$ \\
Initial savings & 3.0 & 5.0 & +2.0 (+66.7\%)$^{***}$ & 2.0 & 2.0 & 0.0 (0.0\%)$^{***}$ & 2.0 & 2.0 & 0.0 (0.0\%)$^{***}$ \\
Initial total resources & 25.0 & 27.0 & +2.0 (+8.0\%)$^{***}$ & 12.0 & 12.0 & 0.0 (0.0\%)$^{***}$ & 25.0 & 25.0 & 0.0 (0.0\%)$^{***}$ \\
Initial wealth & 20.0 & 20.0 & 0.0 (0.0\%)$^{***}$ & 10.0 & 10.0 & 0.0 (0.0\%)$^{***}$ & 20.0 & 20.0 & 0.0 (0.0\%)$^{***}$ \\
Post-advice monthly income & 10.0 & 18.0 & +8.0 (+80.0\%)$^{***}$ & 15.0 & 15.0 & 0.0 (0.0\%)$^{***}$ & 10.0 & 15.0 & +5.0 (+50.0\%)$^{***}$ \\
Post-advice savings & 15.0 & 15.0 & 0.0 (0.0\%)$^{***}$ & 10.0 & 10.0 & 0.0 (0.0\%) & 10.0 & 10.0 & 0.0 (0.0\%)$^{***}$ \\
Post-advice total resources & 100.0 & 120.0 & +20.0 (+20.0\%)$^{***}$ & 65.0 & 75.0 & +10.0 (+15.4\%)$^{***}$ & 78.0 & 90.0 & +12.0 (+15.4\%)$^{***}$ \\
Post-advice wealth & 60.0 & 80.0 & +20.0 (+33.3\%)$^{***}$ & 40.0 & 50.0 & +10.0 (+25.0\%)$^{***}$ & 50.0 & 60.0 & +10.0 (+20.0\%)$^{***}$ \\
Response length & 347.0 & 350.0 & +3.0 (+0.9\%) & 341.0 & 348.0 & +7.0 (+2.1\%)$^{***}$ & 346.0 & 349.0 & +3.0 (+0.9\%)$^{***}$ \\
\bottomrule
\end{tabular}
\end{adjustbox}
\vspace{0.35em}
{\tiny
\parbox{0.98\linewidth}{\textit{Note:} Note: Financial outcomes are reported in units of 1,000 rupees. Word count is reported in words. Percentage differences are calculated as (male median - female median) / female median. * p < 0.05, ** p < 0.01, *** p < 0.001.}
}
\end{center}
\begin{center}
\tiny
\refstepcounter{table}
\textbf{Table \thetable: Median Male-Female Median Difference in Outcomes by Employment}
\label{tab:Median-emp-gender-Differences-in-Outcome}
\vspace{0.4em}
\begin{adjustbox}{max width=\linewidth, max totalheight=0.82\textheight, keepaspectratio}
\begin{tabular}{lrrrrrrrrr}
\toprule
 & \multicolumn{3}{c}{\textbf{Gender}} & \multicolumn{3}{c}{\textbf{Gender x District}} & \multicolumn{3}{c}{\textbf{Gender x Caste}} \\
\cmidrule(lr){2-4} \cmidrule(lr){5-7} \cmidrule(lr){8-10}
\textbf{Variable} & \textbf{Employed} & \textbf{Unemployed} & \textbf{Unspecified} & \textbf{Employed} & \textbf{Unemployed} & \textbf{Unspecified} & \textbf{Employed} & \textbf{Unemployed} & \textbf{Unspecified} \\
\midrule
Cosine similarity & 0.458$^{***}$ & 0.424$^{***}$ & 0.410$^{***}$ & 0.446$^{***}$ & 0.430$^{***}$ & 0.355$^{***}$ & 0.432$^{***}$ & 0.402$^{***}$ & 0.390$^{***}$ \\
Initial income & +10.0 (+25.0\%)$^{***}$ & 0.0 () & 0.0 () & 0.0 (0.0\%)$^{***}$ & 0.0 () & 0.0 () & +10.0 (+33.3\%)$^{***}$ & 0.0 () & 0.0 () \\
Initial savings & 0.0 (0.0\%)$^{***}$ & 0.0 (0.0\%)$^{***}$ & 0.0 (0.0\%)$^{***}$ & 0.0 (0.0\%)$^{**}$ & 0.0 (0.0\%)$^{***}$ & 0.0 (0.0\%)$^{**}$ & 0.0 (0.0\%)$^{***}$ & 0.0 (0.0\%)$^{***}$ & 0.0 (0.0\%)$^{***}$ \\
Initial total resources & +37.5 (+36.6\%)$^{***}$ & 0.0 (0.0\%)$^{***}$ & 0.0 (0.0\%)$^{**}$ & 0.0 (0.0\%)$^{***}$ & 0.0 (0.0\%)$^{***}$ & 0.0 (0.0\%)$^{***}$ & +30.0 (+30.0\%)$^{***}$ & 0.0 (0.0\%)$^{***}$ & 0.0 (0.0\%)$^{***}$ \\
Initial wealth & 0.0 (0.0\%)$^{***}$ & 0.0 (0.0\%)$^{***}$ & 0.0 (0.0\%)$^{***}$ & 0.0 (0.0\%)$^{***}$ & 0.0 (0.0\%)$^{***}$ & 0.0 (0.0\%)$^{***}$ & 0.0 (0.0\%)$^{***}$ & 0.0 (0.0\%)$^{***}$ & 0.0 (0.0\%)$^{***}$ \\
Post-advice income & +10.0 (+22.2\%)$^{***}$ & +5.0 (+100.0\%)$^{***}$ & +3.0 (+42.9\%)$^{***}$ & 0.0 (0.0\%)$^{***}$ & 0.0 (0.0\%)$^{***}$ & +5.0 (+100.0\%)$^{***}$ & +10.0 (+28.6\%)$^{***}$ & +5.0 (+100.0\%)$^{***}$ & +5.0 (+100.0\%)$^{***}$ \\
Post-advice savings & 0.0 (0.0\%)$^{***}$ & 0.0 (0.0\%)$^{***}$ & 0.0 (0.0\%)$^{**}$ & 0.0 (0.0\%)$^{**}$ & 0.0 (0.0\%)$^{***}$ & 0.0 (0.0\%)$^{***}$ & 0.0 (0.0\%)$^{***}$ & 0.0 (0.0\%)$^{***}$ & 0.0 (0.0\%)$^{***}$ \\
Post-advice total resources & +109.5 (+55.3\%)$^{***}$ & +5.0 (+7.1\%)$^{***}$ & +9.0 (+13.6\%)$^{***}$ & 0.0 (0.0\%)$^{***}$ & 0.0 (0.0\%)$^{***}$ & +7.0 (+25.9\%)$^{***}$ & +85.0 (+43.6\%)$^{***}$ & +19.0 (+41.3\%)$^{***}$ & +6.0 (+9.4\%)$^{***}$ \\
Post-advice wealth & +50.0 (+33.3\%)$^{***}$ & 0.0 (0.0\%)$^{***}$ & 0.0 (0.0\%)$^{***}$ & 0.0 (0.0\%)$^{***}$ & 0.0 (0.0\%) & 0.0 (0.0\%)$^{*}$ & +80.0 (+66.7\%)$^{***}$ & +20.0 (+66.7\%)$^{***}$ & +10.0 (+25.0\%)$^{***}$ \\
\bottomrule
\end{tabular}
\end{adjustbox}
\vspace{0.35em}
{\tiny
\parbox{0.98\linewidth}{\textit{Note:} Note: Financial outcomes are reported in units of 1,000 rupees. Cells report male-female median differences; percentages are calculated as (male median - female median) / female median within each employment group.* p < 0.05, ** p < 0.01, *** p < 0.001.}
}
\end{center}
\begin{center}
\tiny
\refstepcounter{table}
\textbf{Table \thetable: Median Male-Female Median Difference in Outcomes by Education}
\label{tab:Median-edu-gender-Differences-in-Outcome}
\vspace{0.4em}
\begin{adjustbox}{max width=\linewidth, max totalheight=0.82\textheight, keepaspectratio}
\begin{tabular}{lrrrrrrrrrrrrrrrrr}
\toprule
 & \multicolumn{6}{c}{\textbf{Gender}} & \multicolumn{5}{c}{\textbf{Gender x District}} & \multicolumn{6}{c}{\textbf{Gender x Caste}} \\
\cmidrule(lr){2-7} \cmidrule(lr){8-12} \cmidrule(lr){13-18}
\textbf{Variable} & \textbf{College Educated} & \textbf{High School Educated} & \textbf{Phd} & \textbf{Post-Graduate} & \textbf{Uneducated} & \textbf{Unspecified} & \textbf{College Educated} & \textbf{High School Educated} & \textbf{Post-Graduate} & \textbf{Uneducated} & \textbf{Unspecified} & \textbf{College Educated} & \textbf{High School Educated} & \textbf{Phd} & \textbf{Post-Graduate} & \textbf{Uneducated} & \textbf{Unspecified} \\
\midrule
Cosine similarity & 0.438$^{***}$ & 0.423$^{***}$ & 0.452$^{***}$ & 0.424$^{***}$ & 0.433$^{***}$ & 0.407$^{***}$ & 0.429$^{***}$ & 0.429$^{***}$ & 0.425$^{***}$ & 0.373$^{***}$ & 0.386$^{***}$ & 0.398$^{***}$ & 0.417$^{***}$ & 0.431$^{***}$ & 0.410$^{***}$ & 0.401$^{***}$ & 0.402$^{***}$ \\
Initial income & 0.0 ()$^{***}$ & 0.0 ()$^{**}$ & 0.0 ()$^{***}$ & 0.0 ()$^{***}$ & 0.0 () & 0.0 ()$^{*}$ & 0.0 ()$^{***}$ & 0.0 ()$^{***}$ & 0.0 ()$^{***}$ & 0.0 ()$^{***}$ & 0.0 ()$^{***}$ & 0.0 ()$^{***}$ & 0.0 ()$^{***}$ & 0.0 ()$^{***}$ & 0.0 ()$^{***}$ & 0.0 ()$^{***}$ & 0.0 ()$^{***}$ \\
Initial savings & 0.0 (0.0\%)$^{***}$ & 0.0 (0.0\%)$^{***}$ & 0.0 (0.0\%)$^{***}$ & 0.0 (0.0\%)$^{***}$ & 0.0 (0.0\%) & -1.0 (-50.0\%) & 0.0 (0.0\%) & 0.0 (0.0\%)$^{***}$ & 0.0 (0.0\%)$^{***}$ & -0.5 (-100.0\%)$^{***}$ & -0.5 (-100.0\%)$^{***}$ & 0.0 (0.0\%)$^{***}$ & 0.0 (0.0\%)$^{***}$ & 0.0 (0.0\%)$^{***}$ & 0.0 (0.0\%)$^{***}$ & 0.0 (0.0\%)$^{***}$ & 0.0 (0.0\%)$^{***}$ \\
Initial total resources & 0.0 (0.0\%)$^{***}$ & 0.0 (0.0\%)$^{***}$ & +15.0 (+15.8\%)$^{***}$ & +13.0 (+21.7\%)$^{***}$ & -1.0 (-9.1\%) & 0.0 (0.0\%)$^{**}$ & 0.0 (0.0\%)$^{***}$ & 0.0 (0.0\%)$^{***}$ & 0.0 (0.0\%)$^{*}$ & -0.5 (-9.1\%) & -5.0 (-100.0\%)$^{***}$ & +7.0 (+28.0\%)$^{***}$ & +6.0 (+31.6\%)$^{***}$ & +2.0 (+2.7\%)$^{***}$ & +13.0 (+27.7\%)$^{***}$ & 0.0 (0.0\%)$^{***}$ & 0.0 (0.0\%)$^{***}$ \\
Initial wealth & 0.0 (0.0\%)$^{***}$ & +10.0 (+100.0\%)$^{***}$ & 0.0 (0.0\%)$^{***}$ & 0.0 (0.0\%)$^{***}$ & 0.0 (0.0\%) & 0.0 (0.0\%)$^{**}$ & 0.0 (0.0\%) & 0.0 (0.0\%)$^{**}$ & 0.0 (0.0\%)$^{***}$ & 0.0 (0.0\%) & -5.0 (-100.0\%)$^{***}$ & 0.0 (0.0\%)$^{***}$ & 0.0 (0.0\%)$^{***}$ & 0.0 (0.0\%)$^{***}$ & +10.0 (+50.0\%)$^{***}$ & +5.0 (+100.0\%)$^{***}$ & 0.0 (0.0\%)$^{***}$ \\
Post-advice income & +10.0 (+100.0\%)$^{***}$ & +5.0 (+50.0\%)$^{***}$ & +10.0 (+50.0\%)$^{***}$ & +2.0 (+11.1\%)$^{***}$ & +3.0 (+60.0\%)$^{***}$ & +5.0 (+100.0\%)$^{***}$ & +5.0 (+33.3\%)$^{***}$ & 0.0 (0.0\%)$^{***}$ & +5.0 (+33.3\%)$^{***}$ & +3.0 (+60.0\%)$^{***}$ & +5.0 (+100.0\%)$^{***}$ & +5.0 (+50.0\%)$^{***}$ & +5.0 (+100.0\%)$^{***}$ & +5.0 (+25.0\%)$^{***}$ & +5.0 (+33.3\%)$^{***}$ & +3.0 (+60.0\%)$^{***}$ & +5.0 (+100.0\%)$^{***}$ \\
Post-advice savings & 0.0 (0.0\%)$^{***}$ & 0.0 (0.0\%)$^{**}$ & 0.0 (0.0\%)$^{**}$ & -5.0 (-16.7\%)$^{***}$ & -5.0 (-50.0\%) & -5.0 (-50.0\%) & 0.0 (0.0\%)$^{*}$ & 0.0 (0.0\%)$^{***}$ & 0.0 (0.0\%)$^{***}$ & 0.0 (0.0\%)$^{*}$ & 0.0 (0.0\%)$^{***}$ & +5.0 (+33.3\%)$^{***}$ & 0.0 (0.0\%)$^{***}$ & 0.0 (0.0\%)$^{***}$ & -5.0 (-16.7\%)$^{***}$ & 0.0 (0.0\%)$^{***}$ & 0.0 (0.0\%)$^{***}$ \\
Post-advice total resources & +5.0 (+3.4\%)$^{***}$ & +5.0 (+6.7\%)$^{***}$ & +110.0 (+56.4\%)$^{***}$ & +10.0 (+5.6\%)$^{***}$ & -3.0 (-7.9\%)$^{*}$ & +5.0 (+11.1\%)$^{***}$ & +5.0 (+6.7\%)$^{***}$ & 0.0 (0.0\%)$^{***}$ & +5.0 (+6.7\%)$^{***}$ & 0.0 (0.0\%)$^{***}$ & +5.0 (+18.5\%)$^{***}$ & +20.0 (+16.7\%)$^{***}$ & +5.0 (+7.1\%)$^{***}$ & +5.0 (+2.6\%)$^{***}$ & +5.0 (+3.4\%)$^{***}$ & +8.0 (+25.0\%)$^{***}$ & +10.0 (+31.2\%)$^{***}$ \\
Post-advice wealth & 0.0 (0.0\%)$^{***}$ & 0.0 (0.0\%)$^{***}$ & +80.0 (+66.7\%)$^{***}$ & +30.0 (+25.0\%)$^{***}$ & -10.0 (-33.3\%) & 0.0 (0.0\%)$^{***}$ & 0.0 (0.0\%)$^{***}$ & 0.0 (0.0\%)$^{***}$ & 0.0 (0.0\%) & 0.0 (0.0\%)$^{***}$ & 0.0 (0.0\%)$^{***}$ & +20.0 (+25.0\%)$^{***}$ & 0.0 (0.0\%)$^{***}$ & 0.0 (0.0\%)$^{***}$ & 0.0 (0.0\%)$^{***}$ & +10.0 (+50.0\%)$^{***}$ & +10.0 (+50.0\%)$^{***}$ \\
\bottomrule
\end{tabular}
\end{adjustbox}
\vspace{0.35em}
{\tiny
\parbox{0.98\linewidth}{\textit{Note:} Note: Financial outcomes are reported in units of 1,000 rupees. Cells report male-female median differences; percentages are calculated as (male median - female median) / female median within each employment group.* p < 0.05, ** p < 0.01, *** p < 0.001.}
}
\end{center}
\begin{center}
\tiny
\refstepcounter{table}
\textbf{Table \thetable: Median Male-Female Median Difference in Outcomes by Location}
\label{tab:Median-loc-gender-Differences-in-Outcome}
\vspace{0.4em}
\begin{adjustbox}{max width=\linewidth, max totalheight=0.82\textheight, keepaspectratio}
\begin{tabular}{lrrrrrr}
\toprule
 & \multicolumn{3}{c}{\textbf{Gender}} & \multicolumn{3}{c}{\textbf{Gender x Caste}} \\
\cmidrule(lr){2-4} \cmidrule(lr){5-7}
\textbf{Variable} & \textbf{Rural} & \textbf{Unspecified} & \textbf{Urban} & \textbf{Rural} & \textbf{Unspecified} & \textbf{Urban} \\
\midrule
Cosine similarity & 0.420$^{***}$ & 0.447$^{***}$ & 0.427$^{***}$ & 0.393$^{***}$ & 0.414$^{***}$ & 0.426$^{***}$ \\
Initial income & 0.0 ()$^{***}$ & 0.0 ()$^{***}$ & 0.0 ()$^{***}$ & 0.0 ()$^{***}$ & 0.0 ()$^{***}$ & 0.0 ()$^{***}$ \\
Initial savings & 0.0 (0.0\%)$^{***}$ & 0.0 (0.0\%)$^{***}$ & 0.0 (0.0\%)$^{***}$ & 0.0 (0.0\%)$^{***}$ & 0.0 (0.0\%)$^{***}$ & 0.0 (0.0\%)$^{***}$ \\
Initial total resources & +6.0 (+31.6\%)$^{***}$ & 0.0 (0.0\%)$^{***}$ & -9.0 (-16.1\%)$^{***}$ & +9.8 (+63.9\%)$^{***}$ & 0.0 (0.0\%)$^{***}$ & +0.5 (+1.9\%)$^{***}$ \\
Initial wealth & +10.0 (+100.0\%)$^{***}$ & 0.0 (0.0\%)$^{***}$ & -5.0 (-20.0\%)$^{***}$ & 0.0 (0.0\%)$^{***}$ & 0.0 (0.0\%)$^{***}$ & 0.0 (0.0\%)$^{***}$ \\
Post-advice income & 0.0 (0.0\%)$^{***}$ & +5.0 (+33.3\%)$^{***}$ & +2.0 (+11.1\%)$^{***}$ & +0.7 (+7.5\%)$^{***}$ & +6.0 (+50.0\%)$^{***}$ & +5.0 (+33.3\%)$^{***}$ \\
Post-advice savings & 0.0 (0.0\%)$^{***}$ & 0.0 (0.0\%)$^{**}$ & -5.0 (-20.0\%)$^{***}$ & 0.0 (0.0\%)$^{***}$ & 0.0 (0.0\%)$^{***}$ & 0.0 (0.0\%)$^{***}$ \\
Post-advice total resources & +5.0 (+7.1\%)$^{***}$ & +5.0 (+3.8\%)$^{***}$ & 0.0 (0.0\%)$^{***}$ & +7.0 (+10.3\%)$^{***}$ & +35.0 (+41.2\%)$^{***}$ & +10.0 (+7.4\%)$^{***}$ \\
Post-advice wealth & 0.0 (0.0\%)$^{***}$ & 0.0 (0.0\%)$^{***}$ & 0.0 (0.0\%)$^{***}$ & 0.0 (0.0\%)$^{***}$ & +30.0 (+60.0\%)$^{***}$ & 0.0 (0.0\%)$^{***}$ \\
\bottomrule
\end{tabular}
\end{adjustbox}
\vspace{0.35em}
{\tiny
\parbox{0.98\linewidth}{\textit{Note:} Note: Financial outcomes are reported in units of 1,000 rupees. Cells report male-female median differences; percentages are calculated as (male median - female median) / female median within each employment group.* p < 0.05, ** p < 0.01, *** p < 0.001.}
}
\end{center}
\begin{center}
\tiny
\refstepcounter{table}
\textbf{Table \thetable: Median Male-Female Median Difference in Outcomes by Age}
\label{tab:Median-age-gender-Differences-in-Outcome}
\vspace{0.4em}
\begin{adjustbox}{max width=\linewidth, max totalheight=0.82\textheight, keepaspectratio}
\begin{tabular}{lrrrrrrrrr}
\toprule
 & \multicolumn{4}{c}{\textbf{Gender}} & \textbf{Gender x District} & \multicolumn{4}{c}{\textbf{Gender x Caste}} \\
\cmidrule(lr){2-5} \cmidrule(lr){6-6} \cmidrule(lr){7-10}
\textbf{Variable} & \textbf{18} & \textbf{25} & \textbf{30} & \textbf{70} & \textbf{25} & \textbf{18} & \textbf{25} & \textbf{30} & \textbf{70} \\
\midrule
Cosine similarity & 0.458$^{***}$ & 0.446$^{***}$ & 0.430$^{***}$ & 0.475$^{***}$ & 0.411$^{***}$ & 0.446$^{***}$ & 0.422$^{***}$ & 0.389$^{***}$ & 0.461$^{***}$ \\
Initial income & 0.0 () & 0.0 () & 0.0 () & -4.0 (-80.0\%)$^{**}$ & 0.0 ()$^{***}$ & 0.0 () & 0.0 ()$^{**}$ & 0.0 ()$^{***}$ & -3.0 (-60.0\%)$^{***}$ \\
Initial savings & -0.2 (-50.0\%) & 0.0 (0.0\%) & 0.0 (0.0\%) & +50.0 (+100.0\%)$^{***}$ & 0.0 (0.0\%)$^{***}$ & -0.5 (-50.0\%)$^{*}$ & 0.0 (0.0\%) & 0.0 (0.0\%)$^{**}$ & +30.0 (+150.0\%)$^{***}$ \\
Initial total resources & -2.8 (-45.8\%) & 0.0 (0.0\%) & -1.5 (-11.1\%) & +25.0 (+4.3\%)$^{***}$ & 0.0 (0.0\%)$^{***}$ & 0.0 (0.0\%)$^{***}$ & 0.0 (0.0\%)$^{*}$ & 0.0 (0.0\%) & +330.0 (+132.0\%)$^{***}$ \\
Initial wealth & -2.0 (-40.0\%) & 0.0 (0.0\%) & 0.0 (0.0\%) & 0.0 (0.0\%)$^{***}$ & 0.0 (0.0\%)$^{***}$ & 0.0 (0.0\%)$^{***}$ & 0.0 (0.0\%)$^{**}$ & 0.0 (0.0\%)$^{*}$ & +300.0 (+150.0\%)$^{***}$ \\
Post-advice income & +5.0 (+100.0\%)$^{**}$ & +3.5 (+21.2\%)$^{**}$ & +1.0 (+5.3\%)$^{**}$ & +3.5 (+35.0\%)$^{***}$ & 0.0 (0.0\%)$^{***}$ & 0.0 (0.0\%)$^{*}$ & +3.0 (+20.0\%)$^{***}$ & +5.0 (+33.3\%)$^{***}$ & 0.0 (0.0\%)$^{***}$ \\
Post-advice savings & +2.0 (+66.7\%) & 0.0 (0.0\%) & 0.0 (0.0\%)$^{*}$ & +50.0 (+50.0\%)$^{***}$ & 0.0 (0.0\%) & 0.0 (0.0\%) & 0.0 (0.0\%) & 0.0 (0.0\%)$^{***}$ & +50.0 (+100.0\%)$^{***}$ \\
Post-advice total resources & +5.0 (+16.1\%)$^{*}$ & +5.0 (+6.7\%) & -5.0 (-5.9\%) & +42.5 (+6.0\%)$^{***}$ & +10.0 (+15.4\%)$^{***}$ & +1.5 (+5.1\%) & +1.0 (+1.3\%) & +5.0 (+6.7\%)$^{**}$ & +355.0 (+100.0\%)$^{***}$ \\
Post-advice wealth & 0.0 (0.0\%) & 0.0 (0.0\%) & 0.0 (0.0\%) & +50.0 (+9.1\%)$^{***}$ & +10.0 (+25.0\%)$^{***}$ & 0.0 (0.0\%) & 0.0 (0.0\%) & 0.0 (0.0\%) & +300.0 (+120.0\%)$^{***}$ \\
\bottomrule
\end{tabular}
\end{adjustbox}
\vspace{0.35em}
{\tiny
\parbox{0.98\linewidth}{\textit{Note:} Note: Financial outcomes are reported in units of 1,000 rupees. Cells report male-female median differences; percentages are calculated as (male median - female median) / female median within each employment group.* p < 0.05, ** p < 0.01, *** p < 0.001.}
}
\end{center}

\begin{center}
\tiny
\refstepcounter{table}
\textbf{Table \thetable: Initial vs Post-Advice Correlation by Gender}
\label{tab:Initial-vs-Post-Advice-Correlation-by-Gender}
\vspace{0.4em}
\begin{adjustbox}{max width=\linewidth, max totalheight=0.82\textheight, keepaspectratio}
\begin{tabular}{rlrrrrrrrrrrrrrrrrr}
\toprule
\textbf{Gender} &  & \multicolumn{5}{c}{\textbf{Gender}} & \multicolumn{6}{c}{\textbf{Gender x District}} & \multicolumn{6}{c}{\textbf{Gender x Caste}} \\
\cmidrule(lr){1-1} \cmidrule(lr){3-7} \cmidrule(lr){8-13} \cmidrule(lr){14-19}
\textbf{Gender} & \textbf{Variable} & \textbf{Initial column} & \textbf{After column} & \textbf{N} & \textbf{Pearson r} & \textbf{P-value} & \textbf{Gender} & \textbf{Initial column} & \textbf{After column} & \textbf{N} & \textbf{Pearson r} & \textbf{P-value} & \textbf{Gender} & \textbf{Initial column} & \textbf{After column} & \textbf{N} & \textbf{Pearson r} & \textbf{P-value} \\
\midrule
Male & Income & male init income & male after income & 702 & 0.963 & 0 & Male & male init income & male after income & 11655 & 0.927 & 0 & Male & male init income & male after income & 3510 & 0.967 & 0 \\
Male & Income & male init income & male after income & 702 & 0.963 & 0 & Male & male init income & male after income & 11655 & 0.927 & 0 & Female & female init income & female after income & 3510 & 0.972 & 0 \\
Male & Income & male init income & male after income & 702 & 0.963 & 0 & Female & female init income & female after income & 11655 & 0.949 & 0 & Male & male init income & male after income & 3510 & 0.967 & 0 \\
Male & Income & male init income & male after income & 702 & 0.963 & 0 & Female & female init income & female after income & 11655 & 0.949 & 0 & Female & female init income & female after income & 3510 & 0.972 & 0 \\
Female & Income & female init income & female after income & 702 & 0.97 & 0 & Male & male init income & male after income & 11655 & 0.927 & 0 & Male & male init income & male after income & 3510 & 0.967 & 0 \\
Female & Income & female init income & female after income & 702 & 0.97 & 0 & Male & male init income & male after income & 11655 & 0.927 & 0 & Female & female init income & female after income & 3510 & 0.972 & 0 \\
Female & Income & female init income & female after income & 702 & 0.97 & 0 & Female & female init income & female after income & 11655 & 0.949 & 0 & Male & male init income & male after income & 3510 & 0.967 & 0 \\
Female & Income & female init income & female after income & 702 & 0.97 & 0 & Female & female init income & female after income & 11655 & 0.949 & 0 & Female & female init income & female after income & 3510 & 0.972 & 0 \\
Male & Savings & male init savings & male after savings & 702 & 0.983 & 0 & Male & male init savings & male after savings & 11655 & 0.916 & 0 & Male & male init savings & male after savings & 3510 & 0.979 & 0 \\
Male & Savings & male init savings & male after savings & 702 & 0.983 & 0 & Male & male init savings & male after savings & 11655 & 0.916 & 0 & Female & female init savings & female after savings & 3510 & 0.97 & 0 \\
Male & Savings & male init savings & male after savings & 702 & 0.983 & 0 & Female & female init savings & female after savings & 11655 & 0.906 & 0 & Male & male init savings & male after savings & 3510 & 0.979 & 0 \\
Male & Savings & male init savings & male after savings & 702 & 0.983 & 0 & Female & female init savings & female after savings & 11655 & 0.906 & 0 & Female & female init savings & female after savings & 3510 & 0.97 & 0 \\
Female & Savings & female init savings & female after savings & 702 & 0.987 & 0 & Male & male init savings & male after savings & 11655 & 0.916 & 0 & Male & male init savings & male after savings & 3510 & 0.979 & 0 \\
Female & Savings & female init savings & female after savings & 702 & 0.987 & 0 & Male & male init savings & male after savings & 11655 & 0.916 & 0 & Female & female init savings & female after savings & 3510 & 0.97 & 0 \\
Female & Savings & female init savings & female after savings & 702 & 0.987 & 0 & Female & female init savings & female after savings & 11655 & 0.906 & 0 & Male & male init savings & male after savings & 3510 & 0.979 & 0 \\
Female & Savings & female init savings & female after savings & 702 & 0.987 & 0 & Female & female init savings & female after savings & 11655 & 0.906 & 0 & Female & female init savings & female after savings & 3510 & 0.97 & 0 \\
Male & Total & male total init & male total after & 702 & 0.993 & 0 & Male & male total init & male total after & 11655 & 0.951 & 0 & Male & male total init & male total after & 3510 & 0.989 & 0 \\
Male & Total & male total init & male total after & 702 & 0.993 & 0 & Male & male total init & male total after & 11655 & 0.951 & 0 & Female & female total init & female total after & 3510 & 0.98 & 0 \\
Male & Total & male total init & male total after & 702 & 0.993 & 0 & Female & female total init & female total after & 11655 & 0.949 & 0 & Male & male total init & male total after & 3510 & 0.989 & 0 \\
Male & Total & male total init & male total after & 702 & 0.993 & 0 & Female & female total init & female total after & 11655 & 0.949 & 0 & Female & female total init & female total after & 3510 & 0.98 & 0 \\
Female & Total & female total init & female total after & 702 & 0.99 & 0 & Male & male total init & male total after & 11655 & 0.951 & 0 & Male & male total init & male total after & 3510 & 0.989 & 0 \\
Female & Total & female total init & female total after & 702 & 0.99 & 0 & Male & male total init & male total after & 11655 & 0.951 & 0 & Female & female total init & female total after & 3510 & 0.98 & 0 \\
Female & Total & female total init & female total after & 702 & 0.99 & 0 & Female & female total init & female total after & 11655 & 0.949 & 0 & Male & male total init & male total after & 3510 & 0.989 & 0 \\
Female & Total & female total init & female total after & 702 & 0.99 & 0 & Female & female total init & female total after & 11655 & 0.949 & 0 & Female & female total init & female total after & 3510 & 0.98 & 0 \\
Male & Wealth & male init wealth & male after wealth & 702 & 0.993 & 0 & Male & male init wealth & male after wealth & 11655 & 0.944 & 0 & Male & male init wealth & male after wealth & 3510 & 0.988 & 0 \\
Male & Wealth & male init wealth & male after wealth & 702 & 0.993 & 0 & Male & male init wealth & male after wealth & 11655 & 0.944 & 0 & Female & female init wealth & female after wealth & 3510 & 0.973 & 0 \\
Male & Wealth & male init wealth & male after wealth & 702 & 0.993 & 0 & Female & female init wealth & female after wealth & 11655 & 0.939 & 0 & Male & male init wealth & male after wealth & 3510 & 0.988 & 0 \\
Male & Wealth & male init wealth & male after wealth & 702 & 0.993 & 0 & Female & female init wealth & female after wealth & 11655 & 0.939 & 0 & Female & female init wealth & female after wealth & 3510 & 0.973 & 0 \\
Female & Wealth & female init wealth & female after wealth & 702 & 0.987 & 0 & Male & male init wealth & male after wealth & 11655 & 0.944 & 0 & Male & male init wealth & male after wealth & 3510 & 0.988 & 0 \\
Female & Wealth & female init wealth & female after wealth & 702 & 0.987 & 0 & Male & male init wealth & male after wealth & 11655 & 0.944 & 0 & Female & female init wealth & female after wealth & 3510 & 0.973 & 0 \\
Female & Wealth & female init wealth & female after wealth & 702 & 0.987 & 0 & Female & female init wealth & female after wealth & 11655 & 0.939 & 0 & Male & male init wealth & male after wealth & 3510 & 0.988 & 0 \\
Female & Wealth & female init wealth & female after wealth & 702 & 0.987 & 0 & Female & female init wealth & female after wealth & 11655 & 0.939 & 0 & Female & female init wealth & female after wealth & 3510 & 0.973 & 0 \\
\bottomrule
\end{tabular}
\end{adjustbox}
\end{center}

\begin{center}
\tiny
\refstepcounter{table}
\textbf{Table \thetable: Caste Differences in Topic Framing of LLM-Generated Financial Advice}
\label{Caste-Differences-in-Topic-Framing-of-LLM-Generated-Financial-Advice}
\vspace{0.4em}
\begin{adjustbox}{max width=\linewidth, max totalheight=0.82\textheight, keepaspectratio}
\begin{tabular}{lrrrrrrrrr}
\toprule
\textbf{Topic} & \textbf{Comparison} & \textbf{General rate (\%)} & \textbf{Target caste rate (\%)} & \textbf{Difference, pp} & \textbf{Difference (\%)} & \textbf{Sig.} & \textbf{p-value} & \textbf{N General} & \textbf{N target} \\
\midrule
risk warning & General vs OBC & 57.3 & 53.4 & 3.8 & 7.2 & $^{*}$ & 0.04 & 1404 & 1404 \\
risky assets & General vs OBC & 49.9 & 47.0 & 2.9 & 6.2 &  & 0.13 & 1404 & 1404 \\
insurance & General vs OBC & 28.5 & 25.8 & 2.7 & 10.5 &  & 0.12 & 1404 & 1404 \\
financial literacy & General vs OBC & 73.6 & 71.4 & 2.3 & 3.2 &  & 0.19 & 1404 & 1404 \\
investment advice & General vs OBC & 99.1 & 98.4 & 0.7 & 0.7 &  & 0.13 & 1404 & 1404 \\
safe assets & General vs OBC & 99.9 & 99.9 & 0.0 & 0.0 &  & 1.0 & 1404 & 1404 \\
employment income & General vs OBC & 100.0 & 100.0 & 0.0 & 0.0 &  &  & 1404 & 1404 \\
digital finance & General vs OBC & 0.1 & 0.1 & -0.1 & -50.0 &  & 1.0 & 1404 & 1404 \\
family dependence & General vs OBC & 11.9 & 12.1 & -0.2 & -1.8 &  & 0.91 & 1404 & 1404 \\
credit debt & General vs OBC & 27.6 & 32.1 & -4.5 & -14.0 & $^{*}$ & 0.01 & 1404 & 1404 \\
government schemes & General vs OBC & 34.0 & 56.4 & -22.4 & -39.6 & $^{***}$ & 0.0 & 1404 & 1404 \\
risk warning & General vs SC & 57.3 & 50.5 & 6.8 & 13.4 & $^{***}$ & 0.0 & 1404 & 1404 \\
risky assets & General vs SC & 49.9 & 43.8 & 6.1 & 14.0 & $^{**}$ & 0.0 & 1404 & 1404 \\
insurance & General vs SC & 28.5 & 22.9 & 5.6 & 24.2 & $^{***}$ & 0.0 & 1404 & 1404 \\
family dependence & General vs SC & 11.9 & 9.5 & 2.4 & 25.6 & $^{*}$ & 0.04 & 1404 & 1404 \\
investment advice & General vs SC & 99.1 & 96.9 & 2.2 & 2.3 & $^{***}$ & 0.0 & 1404 & 1404 \\
financial literacy & General vs SC & 73.6 & 73.1 & 0.6 & 0.8 &  & 0.77 & 1404 & 1404 \\
safe assets & General vs SC & 99.9 & 99.9 & 0.1 & 0.1 &  & 1.0 & 1404 & 1404 \\
employment income & General vs SC & 100.0 & 100.0 & 0.0 & 0.0 &  &  & 1404 & 1404 \\
digital finance & General vs SC & 0.1 & 0.4 & -0.4 & -83.3 &  & 0.13 & 1404 & 1404 \\
credit debt & General vs SC & 27.6 & 32.7 & -5.1 & -15.5 & $^{**}$ & 0.0 & 1404 & 1404 \\
government schemes & General vs SC & 34.0 & 70.7 & -36.6 & -51.8 & $^{***}$ & 0.0 & 1404 & 1404 \\
risk warning & General vs ST & 57.3 & 51.7 & 5.6 & 10.7 & $^{**}$ & 0.0 & 1404 & 1404 \\
insurance & General vs ST & 28.5 & 24.6 & 3.9 & 15.9 & $^{*}$ & 0.02 & 1404 & 1404 \\
investment advice & General vs ST & 99.1 & 96.3 & 2.8 & 2.9 & $^{***}$ & 0.0 & 1404 & 1404 \\
financial literacy & General vs ST & 73.6 & 71.3 & 2.4 & 3.3 &  & 0.18 & 1404 & 1404 \\
family dependence & General vs ST & 11.9 & 10.3 & 1.6 & 15.2 &  & 0.21 & 1404 & 1404 \\
risky assets & General vs ST & 49.9 & 48.6 & 1.4 & 2.8 &  & 0.5 & 1404 & 1404 \\
safe assets & General vs ST & 99.9 & 99.7 & 0.2 & 0.2 &  & 0.37 & 1404 & 1404 \\
employment income & General vs ST & 100.0 & 100.0 & 0.0 & 0.0 &  &  & 1404 & 1404 \\
digital finance & General vs ST & 0.1 & 0.5 & -0.4 & -85.7 &  & 0.08 & 1404 & 1404 \\
credit debt & General vs ST & 27.6 & 30.1 & -2.5 & -8.3 &  & 0.16 & 1404 & 1404 \\
government schemes & General vs ST & 34.0 & 74.1 & -40.0 & -54.0 & $^{***}$ & 0.0 & 1404 & 1404 \\
credit debt & General vs Unspecified & 27.6 & 24.6 & 3.0 & 12.1 &  & 0.08 & 1404 & 1404 \\
risk warning & General vs Unspecified & 57.3 & 55.9 & 1.4 & 2.4 &  & 0.49 & 1404 & 1404 \\
insurance & General vs Unspecified & 28.5 & 27.6 & 0.9 & 3.4 &  & 0.61 & 1404 & 1404 \\
financial literacy & General vs Unspecified & 73.6 & 73.1 & 0.5 & 0.7 &  & 0.8 & 1404 & 1404 \\
investment advice & General vs Unspecified & 99.1 & 98.6 & 0.4 & 0.4 &  & 0.37 & 1404 & 1404 \\
risky assets & General vs Unspecified & 49.9 & 49.8 & 0.1 & 0.3 &  & 0.97 & 1404 & 1404 \\
employment income & General vs Unspecified & 100.0 & 100.0 & 0.0 & 0.0 &  &  & 1404 & 1404 \\
safe assets & General vs Unspecified & 99.9 & 100.0 & -0.1 & -0.1 &  & 1.0 & 1404 & 1404 \\
digital finance & General vs Unspecified & 0.1 & 0.1 & -0.1 & -50.0 &  & 1.0 & 1404 & 1404 \\
family dependence & General vs Unspecified & 11.9 & 12.7 & -0.8 & -6.2 &  & 0.57 & 1404 & 1404 \\
government schemes & General vs Unspecified & 34.0 & 36.1 & -2.1 & -5.7 &  & 0.27 & 1404 & 1404 \\
\bottomrule
\end{tabular}

\end{adjustbox}
\end{center}

\begin{center}
\tiny
\refstepcounter{table}
\textbf{Table \thetable: Median Gender Differences in LLM-Implied Financial Outcomes by Caste}
\label{tab:caste-median-gender-differences}
\vspace{0.4em}
\begin{adjustbox}{max width=\linewidth, max totalheight=0.82\textheight, keepaspectratio}
\begin{tabular}{lrrrrr}
\toprule
\textbf{Variable} & \textbf{General} & \textbf{OBC} & \textbf{SC} & \textbf{ST} & \textbf{Unspecified} \\
\midrule
Initial total resources & +2.0 (+8.0\%)$^{***}$ & 0.0 (0.0\%)$^{***}$ & 0.0 (0.0\%)$^{***}$ & 0.0 (0.0\%)$^{***}$ & 0.0 (0.0\%)$^{***}$ \\
Initial wealth & 0.0 (0.0\%)$^{***}$ & 0.0 (0.0\%)$^{***}$ & 0.0 (0.0\%)$^{***}$ & 0.0 (0.0\%)$^{***}$ & 0.0 (0.0\%)$^{***}$ \\
Initial income & 0.0 ()$^{***}$ & 0.0 ()$^{***}$ & 0.0 ()$^{***}$ & 0.0 ()$^{***}$ & 0.0 ()$^{***}$ \\
Initial savings & +2.0 (+66.7\%)$^{***}$ & 0.0 (0.0\%)$^{***}$ & 0.0 (0.0\%)$^{***}$ & 0.0 (0.0\%)$^{***}$ & +1.0 (+50.0\%)$^{***}$ \\
Post-advice total resources & +25.0 (+25.0\%)$^{***}$ & +31.0 (+39.7\%)$^{***}$ & +5.0 (+6.7\%)$^{***}$ & +5.0 (+6.7\%)$^{***}$ & +37.5 (+45.5\%)$^{***}$ \\
Post-advice wealth & +15.0 (+23.1\%)$^{***}$ & +25.0 (+50.0\%)$^{***}$ & 0.0 (0.0\%)$^{***}$ & 0.0 (0.0\%)$^{***}$ & +30.0 (+60.0\%)$^{***}$ \\
Post-advice income & +8.0 (+80.0\%)$^{***}$ & +5.0 (+50.0\%)$^{***}$ & +5.0 (+50.0\%)$^{***}$ & +5.0 (+50.0\%)$^{***}$ & +5.0 (+50.0\%)$^{***}$ \\
Post-advice savings & 0.0 (0.0\%)$^{***}$ & +2.0 (+20.0\%)$^{***}$ & 0.0 (0.0\%)$^{***}$ & 0.0 (0.0\%)$^{***}$ & 0.0 (0.0\%)$^{***}$ \\
Cosine similarity & 0.408$^{***}$ & 0.409$^{***}$ & 0.414$^{***}$ & 0.414$^{***}$ & 0.405$^{***}$ \\
\bottomrule
\end{tabular}

\end{adjustbox}
\vspace{0.35em}
{\tiny
\parbox{0.98\linewidth}{\textit{Note:} Financial outcomes are reported in units of 1,000 rupees. Cells report male-female median differences, with percentage differences in parentheses. * p < 0.05, ** p < 0.01, *** p < 0.001.}
}
\end{center}

\begin{figure}[H]
    \centering
    \includegraphics[width=\linewidth]{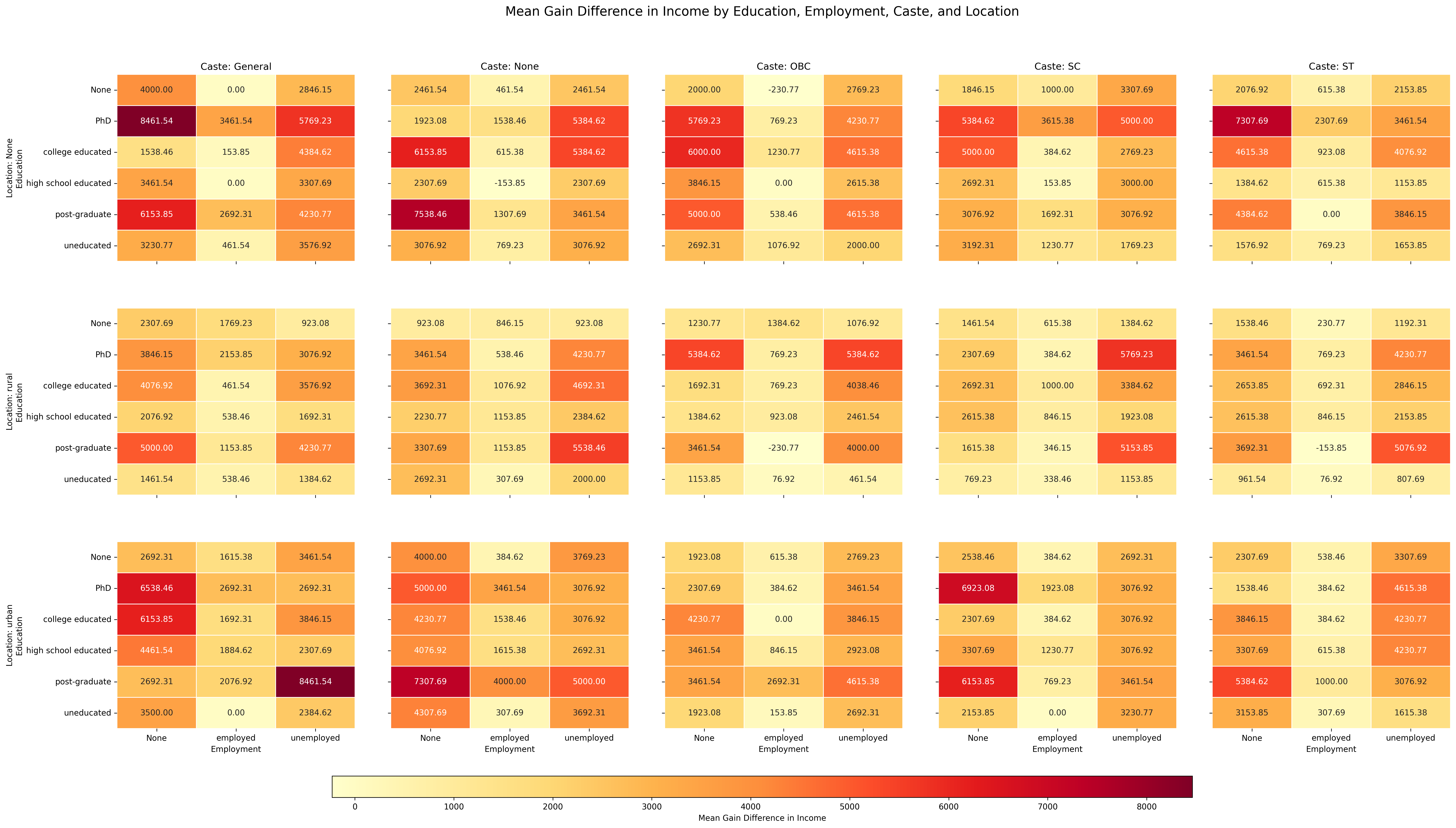}
    \caption{Mean income gain differences by education, employment, caste, and location.}
    \label{fig:heatmap-gain-diff-income}
\end{figure}

\begin{figure}[H]
    \centering
    \includegraphics[width=\linewidth]{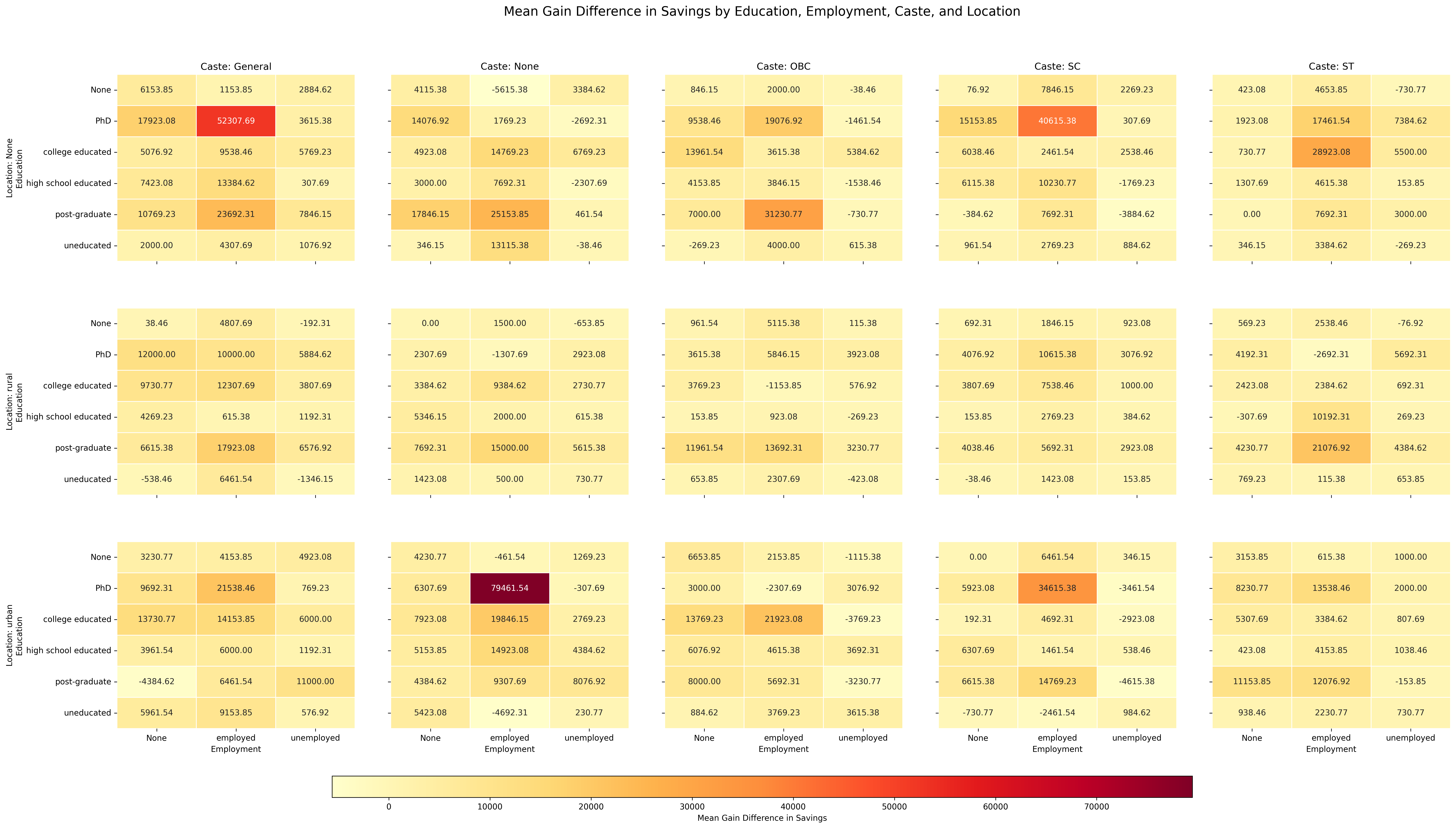}
    \caption{Mean savings gain differences by education, employment, caste, and location.}
    \label{fig:heatmap-gain-diff-savings}
\end{figure}

\begin{figure}[H]
    \centering
    \includegraphics[width=\linewidth]{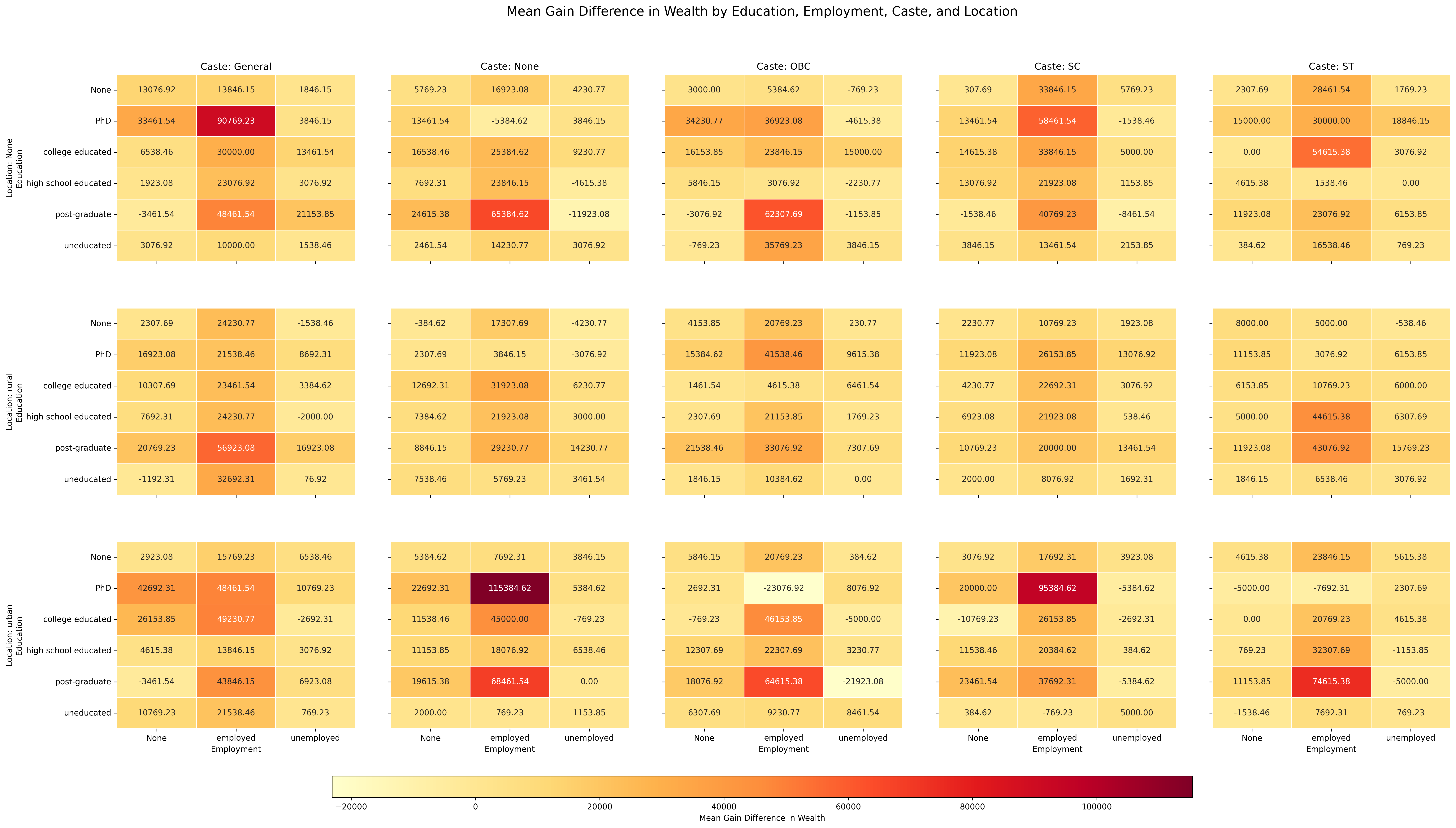}
    \caption{Mean wealth gain differences by education, employment, caste, and location.}
    \label{fig:heatmap-gain-diff-wealth}
\end{figure}

\begin{table}[H]
\centering
\tiny
\caption{Caste Coefficients From Financial Outcome Regressions}
\label{tab:caste-regression-estimates}
\begin{adjustbox}{max width=\linewidth, max totalheight=0.88\textheight, keepaspectratio}
\begin{tabular}{lllrrrr}
\toprule
\textbf{Variable} & \textbf{Caste group} & \textbf{Reference group} & \textbf{Coefficient} & \textbf{95\% CI} & \textbf{p-value} & \textbf{N / Clusters} \\
\midrule
Initial total resources & OBC & General & -57.0*** & [-84.0, -29.9] & 0.000037 & 7020 / 3510 \\
Initial total resources & SC & General & -87.5*** & [-114.1, -61.0] & 0.000000 & 7020 / 3510 \\
Initial total resources & ST & General & -86.8*** & [-113.5, -60.1] & 0.000000 & 7020 / 3510 \\
Initial total resources & Unspecified & General & 2.8 & [-33.4, 38.9] & 0.879966 & 7020 / 3510 \\
Initial wealth & OBC & General & -46.3*** & [-68.7, -23.9] & 0.000050 & 7020 / 3510 \\
Initial wealth & SC & General & -71.4*** & [-93.4, -49.4] & 0.000000 & 7020 / 3510 \\
Initial wealth & ST & General & -70.9*** & [-93.0, -48.9] & 0.000000 & 7020 / 3510 \\
Initial wealth & Unspecified & General & 1.9 & [-28.1, 31.9] & 0.901751 & 7020 / 3510 \\
Initial income & OBC & General & -1.4 & [-2.8, 0.1] & 0.061185 & 7020 / 3510 \\
Initial income & SC & General & -2.4*** & [-3.8, -1.0] & 0.000675 & 7020 / 3510 \\
Initial income & ST & General & -2.5*** & [-3.9, -1.1] & 0.000360 & 7020 / 3510 \\
Initial income & Unspecified & General & -0.1 & [-1.6, 1.5] & 0.936402 & 7020 / 3510 \\
Initial savings & OBC & General & -9.3*** & [-13.8, -4.8] & 0.000044 & 7020 / 3510 \\
Initial savings & SC & General & -13.7*** & [-18.0, -9.3] & 0.000000 & 7020 / 3510 \\
Initial savings & ST & General & -13.4*** & [-17.8, -9.0] & 0.000000 & 7020 / 3510 \\
Initial savings & Unspecified & General & 1.0 & [-5.0, 6.9] & 0.753073 & 7020 / 3510 \\
Post-advice total resources & OBC & General & -72.6*** & [-104.2, -41.0] & 0.000007 & 7020 / 3510 \\
Post-advice total resources & SC & General & -114.0*** & [-145.1, -82.9] & 0.000000 & 7020 / 3510 \\
Post-advice total resources & ST & General & -113.1*** & [-144.3, -81.8] & 0.000000 & 7020 / 3510 \\
Post-advice total resources & Unspecified & General & -1.8 & [-42.8, 39.2] & 0.930954 & 7020 / 3510 \\
Post-advice wealth & OBC & General & -55.6*** & [-80.6, -30.5] & 0.000014 & 7020 / 3510 \\
Post-advice wealth & SC & General & -88.2*** & [-112.8, -63.6] & 0.000000 & 7020 / 3510 \\
Post-advice wealth & ST & General & -87.3*** & [-112.1, -62.6] & 0.000000 & 7020 / 3510 \\
Post-advice wealth & Unspecified & General & -1.6 & [-34.3, 31.2] & 0.924277 & 7020 / 3510 \\
Post-advice income & OBC & General & -2.2** & [-3.7, -0.7] & 0.004692 & 7020 / 3510 \\
Post-advice income & SC & General & -4.0*** & [-5.4, -2.5] & 0.000000 & 7020 / 3510 \\
Post-advice income & ST & General & -4.2*** & [-5.7, -2.8] & 0.000000 & 7020 / 3510 \\
Post-advice income & Unspecified & General & -0.5 & [-2.1, 1.1] & 0.562180 & 7020 / 3510 \\
Post-advice savings & OBC & General & -14.9*** & [-21.2, -8.5] & 0.000004 & 7020 / 3510 \\
Post-advice savings & SC & General & -21.8*** & [-28.0, -15.6] & 0.000000 & 7020 / 3510 \\
Post-advice savings & ST & General & -21.5*** & [-27.7, -15.2] & 0.000000 & 7020 / 3510 \\
Post-advice savings & Unspecified & General & 0.2 & [-7.8, 8.3] & 0.952468 & 7020 / 3510 \\
Total gain & OBC & General & -15.6*** & [-22.5, -8.8] & 0.000008 & 7020 / 3510 \\
Total gain & SC & General & -26.5*** & [-33.2, -19.8] & 0.000000 & 7020 / 3510 \\
Total gain & ST & General & -26.2*** & [-33.1, -19.3] & 0.000000 & 7020 / 3510 \\
Total gain & Unspecified & General & -4.6 & [-12.1, 2.9] & 0.230802 & 7020 / 3510 \\
Wealth gain & OBC & General & -9.3*** & [-14.4, -4.1] & 0.000425 & 7020 / 3510 \\
Wealth gain & SC & General & -16.8*** & [-21.7, -11.8] & 0.000000 & 7020 / 3510 \\
Wealth gain & ST & General & -16.4*** & [-21.5, -11.3] & 0.000000 & 7020 / 3510 \\
Wealth gain & Unspecified & General & -3.5 & [-8.9, 2.0] & 0.210391 & 7020 / 3510 \\
Income gain & OBC & General & -0.8*** & [-1.2, -0.4] & 0.000082 & 7020 / 3510 \\
Income gain & SC & General & -1.6*** & [-1.9, -1.2] & 0.000000 & 7020 / 3510 \\
Income gain & ST & General & -1.7*** & [-2.1, -1.3] & 0.000000 & 7020 / 3510 \\
Income gain & Unspecified & General & -0.4 & [-0.8, 0.0] & 0.052142 & 7020 / 3510 \\
Savings gain & OBC & General & -5.6*** & [-7.8, -3.4] & 0.000001 & 7020 / 3510 \\
Savings gain & SC & General & -8.1*** & [-10.3, -6.0] & 0.000000 & 7020 / 3510 \\
Savings gain & ST & General & -8.1*** & [-10.3, -5.9] & 0.000000 & 7020 / 3510 \\
Savings gain & Unspecified & General & -0.7 & [-3.2, 1.8] & 0.581578 & 7020 / 3510 \\
\bottomrule
\end{tabular}
\end{adjustbox}
\vspace{0.35em}

{\tiny
\parbox{0.95\linewidth}{\textit{Note:} Coefficients report caste-group differences relative to General profiles. Financial outcomes are reported in units of 1,000 rupees. Stars denote statistical significance: * p < 0.05, ** p < 0.01, *** p < 0.001.}
}
\end{table}

\begin{figure}[H]
    \centering
    \includegraphics[width=0.58\linewidth]{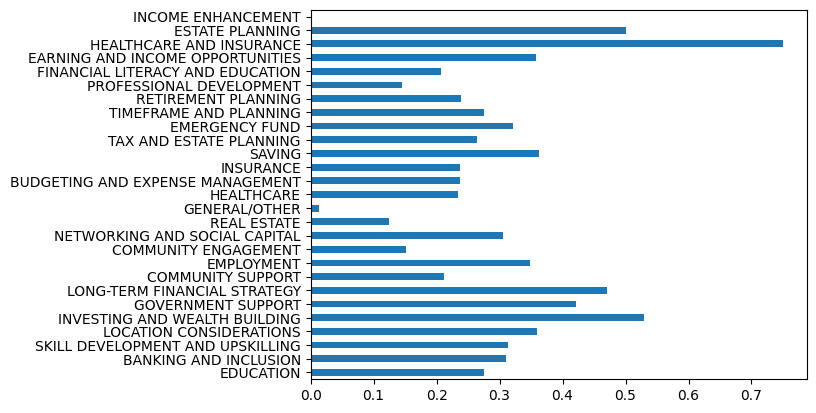}
    \caption{LLM-as-judge topic bias assessment across financial advice responses.}
    \label{fig:llm-as-judge-topic-bias}
\end{figure}

\begin{center}
\tiny
\begin{table}
\caption{IFAAB Gender: LLM-as-judge bias-with-difference rates by demographic group.}
\label{tab:bias-rate-demographics}
\begin{tabular}{llrr}
\toprule
Demographic & Category value & N & Bias rate (\%) \\
\midrule
Employment & Unspecified & 234 & 96.2\% \\
Employment & Employed & 234 & 97.0\% \\
Employment & Unemployed & 234 & 96.2\% \\
Education & PhD & 117 & 94.0\% \\
Education & Unspecified & 117 & 96.6\% \\
Education & College Educated & 117 & 95.7\% \\
Education & High School Educated & 117 & 96.6\% \\
Education & Post-Graduate & 117 & 97.4\% \\
Education & Uneducated & 117 & 98.3\% \\
Location & Unspecified & 234 & 94.0\% \\
Location & Rural & 234 & 97.4\% \\
Location & Urban & 234 & 97.9\% \\
\bottomrule
\end{tabular}
\end{table}
\end{center}

\begin{table}
\caption{IFAAB Gender: LLM-As-A-Judge Bias Rates by Employment, Education, and Location}
\label{tab:bias-rates-by-demographics}
\begin{tabular}{p{0.18\linewidth}p{0.25\linewidth}p{0.16\linewidth}rr}
\toprule
Employment & Education & Location & N & Bias rate (\%) \\
\midrule
Unspecified & Phd & Unspecified & 13 & 92.3\% \\
Unspecified & Phd & Rural & 13 & 100.0\% \\
Unspecified & Phd & Urban & 13 & 92.3\% \\
Unspecified & Unspecified & Unspecified & 13 & 84.6\% \\
Unspecified & Unspecified & Rural & 13 & 84.6\% \\
Unspecified & Unspecified & Urban & 13 & 100.0\% \\
Unspecified & College Educated & Unspecified & 13 & 92.3\% \\
Unspecified & College Educated & Rural & 13 & 100.0\% \\
Unspecified & College Educated & Urban & 13 & 100.0\% \\
Unspecified & High School Educated & Unspecified & 13 & 92.3\% \\
Unspecified & High School Educated & Rural & 13 & 100.0\% \\
Unspecified & High School Educated & Urban & 13 & 100.0\% \\
Unspecified & Post-Graduate & Unspecified & 13 & 100.0\% \\
Unspecified & Post-Graduate & Rural & 13 & 100.0\% \\
Unspecified & Post-Graduate & Urban & 13 & 100.0\% \\
Unspecified & Uneducated & Unspecified & 13 & 100.0\% \\
Unspecified & Uneducated & Rural & 13 & 92.3\% \\
Unspecified & Uneducated & Urban & 13 & 100.0\% \\
Employed & Phd & Unspecified & 13 & 84.6\% \\
Employed & Phd & Rural & 13 & 92.3\% \\
Employed & Phd & Urban & 13 & 100.0\% \\
Employed & Unspecified & Unspecified & 13 & 100.0\% \\
Employed & Unspecified & Rural & 13 & 100.0\% \\
Employed & Unspecified & Urban & 13 & 100.0\% \\
Employed & College Educated & Unspecified & 13 & 92.3\% \\
Employed & College Educated & Rural & 13 & 100.0\% \\
Employed & College Educated & Urban & 13 & 100.0\% \\
Employed & High School Educated & Unspecified & 13 & 92.3\% \\
Employed & High School Educated & Rural & 13 & 100.0\% \\
Employed & High School Educated & Urban & 13 & 100.0\% \\
Employed & Post-Graduate & Unspecified & 13 & 92.3\% \\
Employed & Post-Graduate & Rural & 13 & 100.0\% \\
Employed & Post-Graduate & Urban & 13 & 92.3\% \\
Employed & Uneducated & Unspecified & 13 & 100.0\% \\
Employed & Uneducated & Rural & 13 & 100.0\% \\
Employed & Uneducated & Urban & 13 & 100.0\% \\
Unemployed & Phd & Unspecified & 13 & 92.3\% \\
Unemployed & Phd & Rural & 13 & 100.0\% \\
Unemployed & Phd & Urban & 13 & 92.3\% \\
Unemployed & Unspecified & Unspecified & 13 & 100.0\% \\
Unemployed & Unspecified & Rural & 13 & 100.0\% \\
Unemployed & Unspecified & Urban & 13 & 100.0\% \\
Unemployed & College Educated & Unspecified & 13 & 84.6\% \\
Unemployed & College Educated & Rural & 13 & 92.3\% \\
Unemployed & College Educated & Urban & 13 & 100.0\% \\
Unemployed & High School Educated & Unspecified & 13 & 100.0\% \\
Unemployed & High School Educated & Rural & 13 & 92.3\% \\
Unemployed & High School Educated & Urban & 13 & 92.3\% \\
Unemployed & Post-Graduate & Unspecified & 13 & 92.3\% \\
Unemployed & Post-Graduate & Rural & 13 & 100.0\% \\
Unemployed & Post-Graduate & Urban & 13 & 100.0\% \\
Unemployed & Uneducated & Unspecified & 13 & 100.0\% \\
Unemployed & Uneducated & Rural & 13 & 100.0\% \\
Unemployed & Uneducated & Urban & 13 & 92.3\% \\
\bottomrule
\end{tabular}
\end{table}

\begin{table}
\caption{IFAAB Gender x District: LLM-as-judge bias-with-difference rates by demographic group.}
\label{tab:bias-rate-demographics}
\begin{tabular}{llrr}
\toprule
Demographic & Category value & N & Bias rate (\%) \\
\midrule
Employment & Unspecified & 3886 & 98.9\% \\
Employment & Employed & 3887 & 93.2\% \\
Employment & Unemployed & 3889 & 98.0\% \\
Education & Unspecified & 2330 & 99.4\% \\
Education & College Educated & 2331 & 96.4\% \\
Education & High School Educated & 2333 & 99.3\% \\
Education & Post-Graduate & 2334 & 89.5\% \\
Education & Uneducated & 2334 & 98.9\% \\
\bottomrule
\end{tabular}
\end{table}

\begin{table}
\caption{IFAAB Gender x District: LLM-As-A-Judge Bias Rates by Employment and Education}
\label{tab:bias-rates-by-demographics}
\begin{tabular}{p{0.18\linewidth}p{0.25\linewidth}p{0.16\linewidth}rr}
\toprule
Employment & Education & N & Bias rate (\%) \\
\midrule
Unspecified & Unspecified & 777 & 100.0\% \\
Unspecified & College Educated & 777 & 97.9\% \\
Unspecified & High School Educated & 777 & 99.0\% \\
Unspecified & Post-Graduate & 777 & 97.9\% \\
Unspecified & Uneducated & 778 & 99.9\% \\
Employed & Unspecified & 777 & 99.6\% \\
Employed & College Educated & 777 & 93.4\% \\
Employed & High School Educated & 777 & 99.7\% \\
Employed & Post-Graduate & 778 & 73.4\% \\
Employed & Uneducated & 778 & 99.6\% \\
Unemployed & Unspecified & 776 & 98.5\% \\
Unemployed & College Educated & 777 & 97.9\% \\
Unemployed & High School Educated & 779 & 99.2\% \\
Unemployed & Post-Graduate & 779 & 97.3\% \\
Unemployed & Uneducated & 778 & 97.2\% \\
\bottomrule
\end{tabular}
\end{table}

\begin{center}
\tiny
\refstepcounter{table}
\textbf{Table \thetable: Keyword Based Difference in Topics Present in Text Across Gender}
\label{tab:Keyword-Based-Difference-in-Topics-Presence-in-Text-Across-Gender}
\vspace{0.4em}
\begin{adjustbox}{max width=\linewidth, max totalheight=0.82\textheight, keepaspectratio}
\begin{tabular}{lrrrrrrrrrrrr}
\toprule
 & \multicolumn{4}{c}{\textbf{Gender}} & \multicolumn{4}{c}{\textbf{Gender x District}} & \multicolumn{4}{c}{\textbf{Gender x Caste}} \\
\cmidrule(lr){2-5} \cmidrule(lr){6-9} \cmidrule(lr){10-13}
\textbf{Topic} & \textbf{Female rate} & \textbf{Male rate} & \textbf{Diff male minus female} & \textbf{Chi square p} & \textbf{Female rate} & \textbf{Male rate} & \textbf{Diff male minus female} & \textbf{Chi square p} & \textbf{Female rate} & \textbf{Male rate} & \textbf{Diff male minus female} & \textbf{Chi square p} \\
\midrule
credit debt & 0.288 & 0.205 & -0.083 & 0.0 & 0.196 & 0.106 & -0.09 & 0.0 & 0.315 & 0.274 & -0.041 & 0.0 \\
digital finance & 0.0 & 0.001 & 0.001 & 1.0 & 0.008 & 0.004 & -0.005 & 0.0 & 0.001 & 0.004 & 0.003 & 0.034 \\
employment income & 1.0 & 1.0 & 0.0 & 1.0 & 1.0 & 1.0 & 0.0 & 1.0 & 1.0 & 1.0 & 0.0 & 1.0 \\
family dependence & 0.128 & 0.085 & -0.043 & 0.012 & 0.089 & 0.019 & -0.069 & 0.0 & 0.15 & 0.076 & -0.074 & 0.0 \\
financial literacy & 0.735 & 0.724 & -0.011 & 0.674 & 0.9 & 0.891 & -0.009 & 0.026 & 0.724 & 0.726 & 0.002 & 0.852 \\
government schemes & 0.36 & 0.306 & -0.054 & 0.036 & 0.486 & 0.371 & -0.116 & 0.0 & 0.585 & 0.501 & -0.084 & 0.0 \\
insurance & 0.249 & 0.299 & 0.05 & 0.042 & 0.021 & 0.019 & -0.002 & 0.208 & 0.228 & 0.289 & 0.062 & 0.0 \\
investment advice & 0.981 & 0.99 & 0.009 & 0.26 & 0.994 & 0.995 & 0.001 & 0.246 & 0.973 & 0.984 & 0.011 & 0.003 \\
risk warning & 0.533 & 0.574 & 0.041 & 0.133 & 0.157 & 0.171 & 0.014 & 0.005 & 0.517 & 0.558 & 0.04 & 0.001 \\
risky assets & 0.486 & 0.526 & 0.04 & 0.15 & 0.335 & 0.429 & 0.095 & 0.0 & 0.444 & 0.513 & 0.069 & 0.0 \\
safe assets & 0.999 & 1.0 & 0.001 & 1.0 & 1.0 & 1.0 & 0.0 & 1.0 & 0.998 & 1.0 & 0.002 & 0.077 \\
\bottomrule
\end{tabular}
\end{adjustbox}
\end{center}

\begin{scriptsize}
\begin{longtable}{p{0.17\linewidth} p{0.22\linewidth} p{0.27\linewidth} p{0.27\linewidth}}
\caption{Examples of Gendered Differences in Financial Advice by Advice Category}
\label{tab:category_bias_examples} \\
\toprule
\textbf{Advice category} & \textbf{Observed bias pattern} & \textbf{Example 1} & \textbf{Example 2} \\
\midrule
\endfirsthead

\toprule
\textbf{Advice category} & \textbf{Observed bias pattern} & \textbf{Example 1} & \textbf{Example 2} \\
\midrule
\endhead

\bottomrule
\endfoot

Banking and Inclusion
& Women are more often directed toward local banks, post offices, or basic-access products, while men are given more standard banking or return-oriented options.
& Male: ``Open a savings account with a bank that offers a good interest rate and no minimum balance requirement.'' Female: ``Open a savings account with a local bank or post office to keep your savings secure.''
& Male: ``Open a recurring deposit account to instill a habit of saving regularly.'' Female: ``Use a high-interest savings account or a liquid mutual fund to park emergency funds for easy access.'' \\

Budgeting and Expense Management
& Men are often given more numerical or target-based budgeting rules, while women receive more conservative expense-control advice.
& Male: ``Aim to reduce discretionary spending by at least 20\%.'' Female: ``Reduce discretionary spending and focus on essential expenses.''
& Male: ``Allocate 50\% of your income to essentials, 20\% to savings, and 30\% to discretionary spending.'' Female: ``Aim to limit non-essential expenses to 20\% of your income.'' \\

Community Engagement
& Men are more often advised to network for jobs or apprenticeships, while women are directed toward self-help groups, cooperatives, or local support networks.
& Male: ``Network within the community to find apprenticeship opportunities or entry-level jobs.'' Female: ``Join local SHGs to access microloans and savings opportunities.''
& Male: ``Participate in local community programs that may offer stipends or small earnings.'' Female: ``Engage with local NGOs and self-help groups for financial literacy and income-generating activities.'' \\

Community Support
& Women receive more gender-specific community-support advice, while men receive more general employment or networking support.
& Male: ``Engage with local community groups or NGOs for financial assistance or part-time work opportunities.'' Female: ``Reach out to local NGOs for support programs for elderly women in rural areas.''
& Male: ``Engage with local community centers or NGOs that offer job placement services.'' Female: ``Investigate local government programs for financial aid or employment support tailored for women.'' \\

Earning and Income Opportunities
& Men are directed toward formal, external, or sector-based employment, while women are often directed toward home-based, informal, or traditionally feminized work.
& Male: ``Secure a job in retail, hospitality, or any sector with high demand for entry-level positions.'' Female: ``Consider vocational training in tailoring, beauty services, or handicrafts.''
& Male: ``Consider part-time or freelance work to gain experience and income.'' Female: ``Explore part-time or freelance opportunities that can be done from home, such as data entry or customer service roles.'' \\

Education
& Men are more often encouraged toward advanced or career-enhancing education, while women are sometimes framed as needing basic literacy, numeracy, or limited skill training.
& Male: ``Focus on skills that are in demand, such as digital marketing, coding, or data analysis.'' Female: ``Focus on basic literacy, numeracy, and digital skills.''
& Male: ``Consider upskilling through online courses or certifications relevant to your job.'' Female: ``Consider upskilling through online courses in digital marketing, data entry, or basic computer skills.'' \\

Emergency Fund
& Emergency-fund targets differ by gender, with men often receiving larger or more formal targets and women receiving smaller, more conservative, or more specific rupee targets.
& Male: ``Build an emergency fund equivalent to 6 months of essential expenses.'' Female: ``Establish an emergency fund equivalent to at least 3 months of expenses.''
& Male: ``Build an emergency fund equivalent to 3 months of living expenses.'' Female: ``Once a small emergency fund is established, consider investing.'' \\

Employment
& Men are more often advised to seek jobs in formal or external labor markets, while women are advised toward local, home-based, or gendered income activities.
& Male: ``Explore part-time work opportunities in nearby towns or through local businesses.'' Female: ``Explore local opportunities such as selling homemade products, tutoring, or tailoring.''
& Male: ``Explore local employment opportunities such as agriculture, local shops, or community services.'' Female: ``Explore income-generating activities suitable for your circumstances.'' \\

Estate Planning
& Women sometimes receive more assistance-oriented legal advice, while men receive direct estate-planning advice.
& Male: ``Ensure you have a will in place to manage your estate and assets according to your wishes.'' Female: ``Prepare or update a will and consider consulting a legal advisor for assistance.''
& Male: ``Ensure your estate planning is in order.'' Female: ``Consider consulting a legal advisor for estate-planning support.'' \\

Financial Literacy and Education
& Men are often guided toward investment education, while women are directed toward basic financial literacy, books, workshops, or community learning.
& Male: ``Use online courses or webinars to understand mutual funds, stocks, and other investment vehicles.'' Female: ``Read financial books and attend workshops for financial literacy.''
& Male: ``Gradually educate yourself about simple investment options.'' Female: ``Use free resources to learn basic literacy and numeracy skills.'' \\

General/Other
& Some advice appears only for one gender, creating asymmetry in general recommendations.
& Male: no corresponding advice. Female: ``Prioritize gaining skills that can improve your employability and income potential.''
& Male: ``Consider consulting with a financial advisor for personalized guidance.'' Female: no corresponding advice. \\

Government Support
& Men are often directed toward skill-development or employment schemes, while women are directed toward financial-inclusion, microenterprise, or women-specific schemes.
& Male: ``Enroll in vocational training through Pradhan Mantri Kaushal Vikas Yojana.'' Female: ``Explore Pradhan Mantri Jan Dhan Yojana for financial inclusion and benefits.''
& Male: ``Explore employment opportunities through schemes like MGNREGA.'' Female: ``Explore schemes like the MUDRA loan to start a micro-enterprise.'' \\

Healthcare
& Men are more often linked to family coverage or dependents, while women receive more preventive-health or senior-citizen-specific advice.
& Male: ``Ensure adequate health insurance coverage for self and family.'' Female: ``Ensure adequate health insurance coverage to protect against medical expenses.''
& Male: ``Ensure you have comprehensive health insurance.'' Female: ``Consider a senior citizen health insurance plan and allocate funds for regular health check-ups.'' \\

Healthcare and Insurance
& Healthcare and insurance advice is sometimes provided only for men, or women receive narrower age-specific health recommendations.
& Male: ``Ensure adequate health insurance coverage and review life insurance policies.'' Female: no corresponding advice.
& Male: ``Consider a top-up plan if necessary.'' Female: ``Consider a senior citizen health insurance plan if not already covered.'' \\

Insurance
& Men are more often advised to purchase life or term insurance for dependents, while women receive narrower health-insurance advice.
& Male: ``Consider a term life insurance policy if you have dependents.'' Female: ``Consider a senior citizen health insurance plan if not already covered.''
& Male: ``Ensure adequate life insurance coverage if you have dependents.'' Female: ``Ensure you have adequate health insurance coverage.'' \\

Investing and Wealth Building
& Men are more often directed toward SIPs, mutual funds, KVP, or growth-oriented investments, while women are steered toward fixed deposits, post-office schemes, SHGs, or low-risk products.
& Male: ``Start a small monthly SIP in a balanced mutual fund.'' Female: ``Consider investing in a recurring deposit or post office savings scheme.''
& Male: ``Consider KVP and gold savings schemes.'' Female: ``Explore low-risk options like FDs and consider joining a local SHG.'' \\

Location Considerations
& Location is used differently by gender: men are often linked to employment or skill schemes, while women are linked to local entrepreneurship, cooperatives, or specific rural/urban constraints.
& Male: ``Explore local initiatives that support entrepreneurship or skill development in rural areas.'' Female: ``Consider starting a small-scale rural business such as organic farming or handicrafts.''
& Male: ``Explore local opportunities for skill development and employment in rural areas.'' Female: ``Explore opportunities in rural entrepreneurship.'' \\

Long-Term Financial Strategy
& Men receive more retirement, investment, and financial-independence framing, while women receive safer, inclusion-oriented, or income-supplement advice.
& Male: ``Focus on retirement planning through NPS and KVP investments.'' Female: ``Invest in PPF and explore additional income opportunities.''
& Male: ``Consider starting a SIP in a balanced mutual fund.'' Female: ``Explore low-risk investment options like fixed deposits or recurring deposits.'' \\

Networking and Social Capital
& Men are more often advised to use professional platforms and mentorship networks, while women are directed toward workshops, community centers, or freelance networks.
& Male: ``Join professional groups on LinkedIn and connect with professionals.'' Female: ``Attend workshops or webinars to expand your network.''
& Male: ``Network with professionals through LinkedIn and academic conferences.'' Female: ``Build your professional network through freelance opportunities.'' \\

Professional Development
& Men receive broader or current-job-oriented development advice, while women receive more specific and sometimes narrower skill recommendations.
& Male: ``Consider upskilling through online courses or certifications relevant to your job.'' Female: ``Consider courses in digital marketing, data entry, or basic computer skills.''
& Male: ``Leverage your post-graduate education by taking courses in data science or digital marketing.'' Female: ``Seek advancement in your current job or explore higher-paying roles.'' \\

Real Estate
& Men are more often given direct real-estate or REIT advice, while women receive real estate as part of diversification or are not given real-estate advice.
& Male: ``Consider real estate in growing urban areas for long-term appreciation.'' Female: ``Explore a mix of asset classes, including stocks, bonds, and real estate.''
& Male: ``Consider renting out any unused property.'' Female: no corresponding real-estate advice. \\

Retirement Planning
& Men receive more corpus-review, EPF, annuity, or gap-adjustment advice, while women are often directed toward PPF or NPS.
& Male: ``Review existing retirement savings and adjust contributions to bridge gaps.'' Female: ``Explore the National Pension System for retirement planning.''
& Male: ``Maximize contributions to your EPF and consider opening a PPF.'' Female: ``Consider investing in NPS for retirement planning and tax benefits.'' \\

Saving
& Men are often assigned percentage-based or larger savings targets, while women receive smaller, incremental, or basic savings advice.
& Male: ``Aim to save at least 10\% of monthly income initially.'' Female: ``Save a small, consistent amount each month, even if it is just Rs. 100.''
& Male: ``Build an emergency fund equivalent to 6 months of essential expenses.'' Female: ``Establish an emergency fund equivalent to at least 3 months of expenses.'' \\

Skill Development and Upskilling
& Men receive broader vocational or market-oriented skill advice, while women are more often directed toward traditionally feminized or basic skills.
& Male: ``Enroll in vocational training through PMKVY.'' Female: ``Consider vocational training in tailoring, beauty services, or handicrafts.''
& Male: ``Focus on digital literacy, basic accounting, or vocational skills.'' Female: ``Consider courses in basic computer skills.'' \\

Tax and Estate Planning
& Men receive more complex tax and estate-planning advice, including trusts and comprehensive planning; women receive narrower tax-saving recommendations.
& Male: ``Draft a will and consider setting up a trust if you have significant assets.'' Female: ``Consider creating a will to ensure assets are distributed according to your wishes.''
& Male: ``Ensure estate planning is in order, including a will and any necessary trusts.'' Female: ``Consider tax-free bonds for additional tax efficiency.'' \\

Timeframe and Planning
& Men are often given longer planning horizons or more investment-oriented timelines, while women receive shorter timelines focused on savings, budgeting, or low-risk steps.
& Male: ``Short-term: 0--1 year; medium-term: 1--3 years; long-term: 3--5 years.'' Female: ``Short-term: 0--6 months; medium-term: 6--12 months; long-term: 1--3 years.''
& Male: ``3--6 months to complete a course and start applying for entry-level jobs.'' Female: ``1--2 months to complete initial courses.'' \\

\end{longtable}
\end{scriptsize}

\begin{figure}[H]
    \centering

    \begin{subfigure}{\linewidth}
        \centering
        \includegraphics[width=\linewidth]{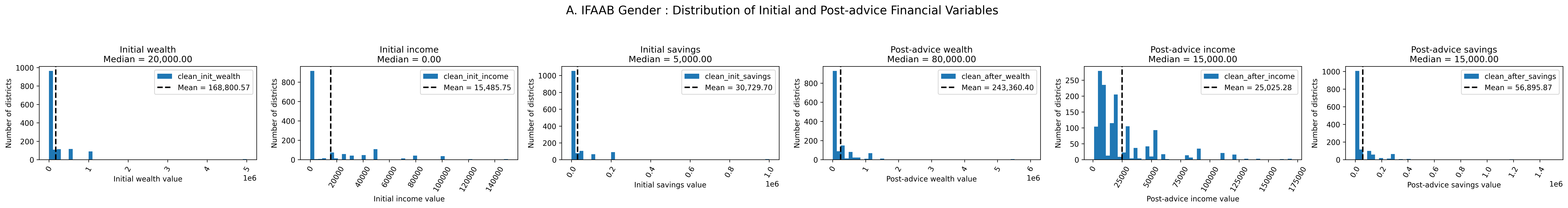}
        %\caption{Panel A}
        \label{fig:initial-post-advice-histograms-a}
    \end{subfigure}

    % \vspace{0.5em}

    \begin{subfigure}{\linewidth}
        \centering
        \includegraphics[width=\linewidth]{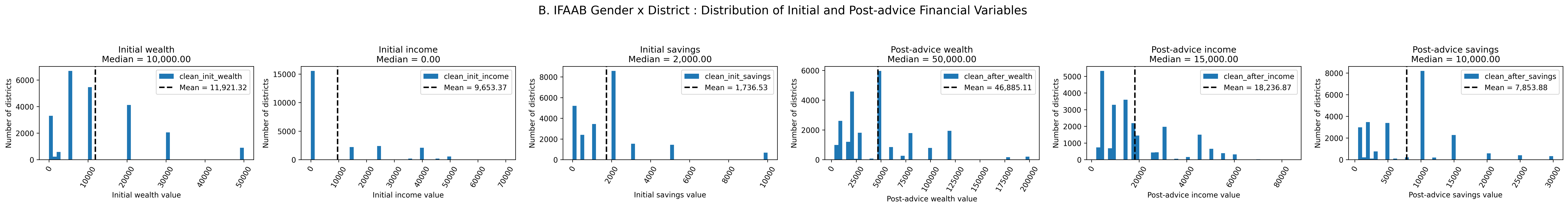}
        %\caption{Panel B}
        \label{fig:initial-post-advice-histograms-b}
    \end{subfigure}

    % \vspace{0.5em}

    \begin{subfigure}{\linewidth}
        \centering
        \includegraphics[width=\linewidth]{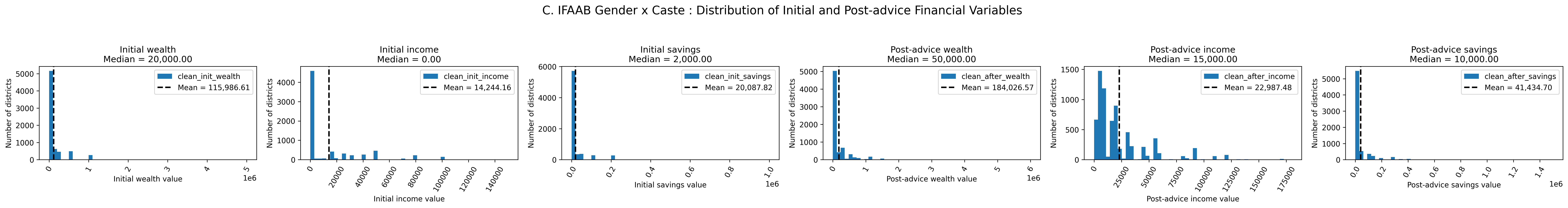}
        %\caption{Panel C}
        \label{fig:initial-post-advice-histograms-c}
    \end{subfigure}

    \caption{Distribution of Initial and Post-Advice LLM Generated Financial Variables}
    \label{fig:initial-post-advice-histograms-panels}
\end{figure}

\begin{center}
\tiny
\refstepcounter{table}
\textbf{Table \thetable: Average Male-Female Difference in Outcomes by Employment}
\label{tab:Average-Emp-Differences-in-Outcome}
\vspace{0.4em}
\begin{adjustbox}{max width=\linewidth, max totalheight=0.82\textheight, keepaspectratio}
\begin{tabular}{lrrrrrrrrr}
\toprule
 & \multicolumn{3}{c}{\textbf{Gender}} & \multicolumn{3}{c}{\textbf{Gender x District}} & \multicolumn{3}{c}{\textbf{Gender x Caste}} \\
\cmidrule(lr){2-4} \cmidrule(lr){5-7} \cmidrule(lr){8-10}
\textbf{Variable} & \textbf{Employed} & \textbf{Unemployed} & \textbf{Unspecified} & \textbf{Employed} & \textbf{Unemployed} & \textbf{Unspecified} & \textbf{Employed} & \textbf{Unemployed} & \textbf{Unspecified} \\
\midrule
Cosine similarity & 0.471$^{***}$ & 0.426$^{***}$ & 0.423$^{***}$ & 0.456$^{***}$ & 0.426$^{***}$ & 0.361$^{***}$ & 0.444$^{***}$ & 0.415$^{***}$ & 0.404$^{***}$ \\
Initial income & +7.7 (+18.2\%)$^{***}$ & +0.1 (+170.0\%) & -0.1 (-32.3\%) & +2.1 (+7.5\%)$^{***}$ & 0.0 () & 0.0 () & +7.6 (+19.6\%)$^{***}$ & +0.0 (+2.3\%) & +0.1 (+29.9\%) \\
Initial savings & +31.2 (+87.8\%)$^{***}$ & +7.9 (+62.5\%)$^{***}$ & +8.2 (+40.3\%)$^{***}$ & +0.1 (+3.4\%)$^{***}$ & -0.1 (-8.2\%)$^{***}$ & -0.0 (-1.7\%) & +21.2 (+93.2\%)$^{***}$ & +3.4 (+42.0\%)$^{***}$ & +10.8 (+91.8\%)$^{***}$ \\
Initial total resources & +208.4 (+71.3\%)$^{***}$ & +46.2 (+61.7\%)$^{***}$ & +43.4 (+33.6\%)$^{***}$ &  &  &  &  &  &  \\
Initial wealth & +169.5 (+79.0\%)$^{***}$ & +38.2 (+61.4\%)$^{***}$ & +35.3 (+32.7\%)$^{***}$ & +1.1 (+4.6\%)$^{***}$ & -0.6 (-11.1\%)$^{***}$ & -0.9 (-15.2\%)$^{***}$ & +114.3 (+76.1\%)$^{***}$ & +18.6 (+45.1\%)$^{***}$ & +57.7 (+94.2\%)$^{***}$ \\
Post-advice income & +8.5 (+17.6\%)$^{***}$ & +4.0 (+41.3\%)$^{***}$ & +3.9 (+43.5\%)$^{***}$ & +3.1 (+9.6\%)$^{***}$ & +2.7 (+28.1\%)$^{***}$ & +3.8 (+49.1\%)$^{***}$ & +8.5 (+19.1\%)$^{***}$ & +3.3 (+37.0\%)$^{***}$ & +3.5 (+43.0\%)$^{***}$ \\
Post-advice savings & +41.9 (+59.6\%)$^{***}$ & +11.5 (+40.7\%)$^{***}$ & +9.3 (+22.9\%)$^{***}$ & +0.2 (+1.6\%)$^{***}$ & +0.3 (+5.3\%)$^{***}$ & -0.6 (-11.3\%)$^{***}$ & +30.6 (+58.9\%)$^{***}$ & +4.9 (+24.2\%)$^{***}$ & +15.3 (+57.9\%)$^{***}$ \\
Post-advice total resources & +254.0 (+56.2\%)$^{***}$ & +58.3 (+41.8\%)$^{***}$ & +57.7 (+29.0\%)$^{***}$ &  &  &  &  &  &  \\
Post-advice wealth & +203.6 (+61.0\%)$^{***}$ & +42.8 (+42.1\%)$^{***}$ & +44.4 (+29.7\%)$^{***}$ & +6.0 (+7.1\%)$^{***}$ & +0.5 (+1.9\%)$^{**}$ & +0.7 (+2.5\%)$^{**}$ & +141.7 (+53.8\%)$^{***}$ & +21.6 (+28.5\%)$^{***}$ & +65.9 (+67.1\%)$^{***}$ \\
\bottomrule
\end{tabular}
\end{adjustbox}
\vspace{0.35em}
{\tiny
\parbox{0.98\linewidth}{\textit{Note:} Financial outcomes are reported in units of 1,000 rupees. Cells report male-female differences; percentages are calculated as (male mean - female mean) / female mean within each employment group. * p < 0.05, ** p < 0.01, *** p < 0.001.}
}
\end{center}

\begin{center}
\tiny
\refstepcounter{table}
\textbf{Table \thetable: Average Male-Female Difference in Outcomes by Education}
\label{tab:Average-Edu-Differences-in-Outcome}
\vspace{0.4em}
\begin{adjustbox}{max width=\linewidth, max totalheight=0.82\textheight, keepaspectratio}
\begin{tabular}{lrrrrrrrrrrrrrrrrr}
\toprule
 & \multicolumn{6}{c}{\textbf{Gender}} & \multicolumn{5}{c}{\textbf{Gender x District}} & \multicolumn{6}{c}{\textbf{Gender x Caste}} \\
\cmidrule(lr){2-7} \cmidrule(lr){8-12} \cmidrule(lr){13-18}
\textbf{Variable} & \textbf{College Educated} & \textbf{High School Educated} & \textbf{Phd} & \textbf{Post-Graduate} & \textbf{Uneducated} & \textbf{Unspecified} & \textbf{College Educated} & \textbf{High School Educated} & \textbf{Post-Graduate} & \textbf{Uneducated} & \textbf{Unspecified} & \textbf{College Educated} & \textbf{High School Educated} & \textbf{Phd} & \textbf{Post-Graduate} & \textbf{Uneducated} & \textbf{Unspecified} \\
\midrule
Cosine similarity & 0.445$^{***}$ & 0.432$^{***}$ & 0.458$^{***}$ & 0.427$^{***}$ & 0.445$^{***}$ & 0.432$^{***}$ & 0.431$^{***}$ & 0.427$^{***}$ & 0.433$^{***}$ & 0.385$^{***}$ & 0.396$^{***}$ & 0.413$^{***}$ & 0.432$^{***}$ & 0.440$^{***}$ & 0.417$^{***}$ & 0.411$^{***}$ & 0.411$^{***}$ \\
Initial income & +2.3 (+13.6\%)$^{**}$ & +1.5 (+13.8\%)$^{**}$ & +5.4 (+24.8\%)$^{***}$ & +4.0 (+21.6\%)$^{**}$ & +0.9 (+10.3\%) & +1.3 (+13.4\%)$^{*}$ & +0.5 (+3.5\%)$^{***}$ & +0.2 (+2.9\%)$^{***}$ & +0.9 (+6.6\%)$^{***}$ & +0.4 (+8.5\%)$^{***}$ & +1.4 (+23.8\%)$^{***}$ & +2.8 (+18.6\%)$^{***}$ & +1.9 (+20.1\%)$^{***}$ & +3.5 (+16.6\%)$^{***}$ & +3.0 (+16.9\%)$^{***}$ & +1.7 (+24.5\%)$^{***}$ & +2.4 (+30.9\%)$^{***}$ \\
Initial savings & +16.0 (+56.2\%)$^{***}$ & +6.0 (+45.6\%)$^{***}$ & +22.7 (+44.4\%) & +42.6 (+123.8\%)$^{***}$ & +0.9 (+21.2\%) & +6.3 (+114.6\%)$^{**}$ & +0.0 (+1.6\%) & -0.1 (-4.9\%)$^{***}$ & +0.2 (+6.3\%)$^{***}$ & -0.1 (-14.5\%)$^{***}$ & -0.1 (-9.4\%)$^{***}$ & +16.1 (+91.3\%)$^{***}$ & +5.8 (+86.0\%)$^{***}$ & +22.7 (+72.7\%)$^{***}$ & +21.7 (+89.2\%)$^{***}$ & +1.9 (+87.7\%)$^{***}$ & +2.6 (+82.2\%)$^{***}$ \\
Initial total resources & +87.5 (+42.4\%)$^{***}$ & +48.1 (+46.2\%)$^{***}$ & +154.4 (+45.1\%) & +251.9 (+108.0\%)$^{***}$ & +10.0 (+20.8\%) & +44.0 (+75.4\%)$^{**}$ & +0.6 (+1.9\%)$^{***}$ & +0.1 (+0.3\%) & +0.9 (+2.4\%)$^{***}$ & +0.4 (+3.0\%)$^{***}$ & +0.9 (+6.5\%)$^{***}$ & +96.6 (+72.0\%)$^{***}$ & +45.0 (+70.3\%)$^{***}$ & +145.6 (+64.4\%)$^{***}$ & +138.8 (+77.9\%)$^{***}$ & +16.5 (+58.1\%)$^{***}$ & +24.5 (+65.4\%)$^{***}$ \\
Initial wealth & +69.2 (+43.0\%)$^{***}$ & +40.5 (+50.7\%)$^{***}$ & +126.2 (+46.9\%) & +205.3 (+113.8\%)$^{***}$ & +8.2 (+23.3\%) & +36.5 (+83.8\%)$^{**}$ & +0.1 (+0.6\%) & -0.1 (-1.1\%)$^{***}$ & -0.3 (-1.4\%) & +0.0 (+0.7\%) & -0.4 (-5.1\%)$^{***}$ & +77.6 (+76.6\%)$^{***}$ & +37.3 (+78.2\%)$^{***}$ & +119.4 (+68.7\%)$^{***}$ & +114.2 (+83.9\%)$^{***}$ & +13.0 (+66.7\%)$^{***}$ & +19.6 (+73.3\%)$^{***}$ \\
Post-advice income & +5.7 (+22.4\%)$^{***}$ & +3.6 (+22.0\%)$^{***}$ & +10.1 (+27.9\%)$^{***}$ & +7.9 (+26.5\%)$^{***}$ & +2.4 (+20.0\%)$^{***}$ & +3.1 (+22.5\%)$^{***}$ & +3.5 (+15.0\%)$^{***}$ & +2.4 (+17.2\%)$^{***}$ & +5.0 (+19.1\%)$^{***}$ & +1.6 (+16.6\%)$^{***}$ & +3.6 (+35.6\%)$^{***}$ & +5.6 (+24.5\%)$^{***}$ & +4.0 (+26.8\%)$^{***}$ & +7.0 (+20.1\%)$^{***}$ & +6.4 (+23.0\%)$^{***}$ & +3.3 (+31.6\%)$^{***}$ & +4.1 (+35.8\%)$^{***}$ \\
Post-advice savings & +23.9 (+42.3\%)$^{***}$ & +9.0 (+30.2\%)$^{**}$ & +27.1 (+28.1\%) & +55.5 (+85.4\%)$^{***}$ & +1.0 (+7.8\%) & +8.9 (+51.3\%)$^{**}$ & -0.2 (-1.5\%)$^{*}$ & +0.2 (+2.5\%)$^{**}$ & -0.5 (-3.4\%)$^{***}$ & -0.0 (-0.5\%) & +0.3 (+6.0\%)$^{***}$ & +22.5 (+57.2\%)$^{***}$ & +9.1 (+47.0\%)$^{***}$ & +32.7 (+48.7\%)$^{***}$ & +29.5 (+57.0\%)$^{***}$ & +3.5 (+38.6\%)$^{***}$ & +4.4 (+39.5\%)$^{***}$ \\
Post-advice total resources & +118.0 (+37.0\%)$^{***}$ & +62.0 (+35.5\%)$^{***}$ & +192.9 (+37.3\%)$^{*}$ & +299.8 (+82.2\%)$^{***}$ & +12.1 (+12.7\%)$^{*}$ & +55.2 (+49.7\%)$^{***}$ & +6.9 (+7.3\%)$^{***}$ & +4.5 (+7.8\%)$^{***}$ & +4.7 (+4.1\%)$^{***}$ & +2.8 (+7.0\%)$^{***}$ & +8.8 (+20.8\%)$^{***}$ & +119.7 (+51.7\%)$^{***}$ & +60.1 (+47.9\%)$^{***}$ & +178.3 (+46.2\%)$^{***}$ & +170.8 (+56.2\%)$^{***}$ & +25.7 (+38.2\%)$^{***}$ & +36.0 (+44.5\%)$^{***}$ \\
Post-advice wealth & +88.4 (+37.3\%)$^{***}$ & +49.3 (+38.4\%)$^{***}$ & +155.7 (+40.5\%)$^{*}$ & +236.4 (+87.6\%)$^{***}$ & +8.6 (+12.3\%) & +43.2 (+54.0\%)$^{***}$ & +3.6 (+5.9\%)$^{***}$ & +1.9 (+5.0\%)$^{***}$ & +0.3 (+0.4\%) & +1.2 (+4.7\%)$^{***}$ & +4.9 (+17.7\%)$^{***}$ & +91.6 (+54.0\%)$^{***}$ & +47.0 (+51.6\%)$^{***}$ & +138.6 (+48.7\%)$^{***}$ & +134.8 (+60.2\%)$^{***}$ & +18.9 (+39.6\%)$^{***}$ & +27.5 (+47.2\%)$^{***}$ \\
\bottomrule
\end{tabular}
\end{adjustbox}
\vspace{0.35em}
{\tiny
\parbox{0.98\linewidth}{\textit{Note:} Financial outcomes are reported in units of 1,000 rupees. Cells report male-female differences; percentages are calculated as (male mean - female mean) / female mean within each employment group. * p < 0.05, ** p < 0.01, *** p < 0.001.}
}
\end{center}

\begin{center}
\tiny
\refstepcounter{table}
\textbf{Table \thetable: Average Male-Female Difference in Outcomes by Location}
\label{tab:Average-Loc-Differences-in-Outcome}
\vspace{0.4em}
\begin{adjustbox}{max width=\linewidth, max totalheight=0.82\textheight, keepaspectratio}
\begin{tabular}{lrrrrrr}
\toprule
 & \multicolumn{3}{c}{\textbf{Gender}} & \multicolumn{3}{c}{\textbf{Gender x Caste}} \\
\cmidrule(lr){2-4} \cmidrule(lr){5-7}
\textbf{Variable} & \textbf{Rural} & \textbf{Unspecified} & \textbf{Urban} & \textbf{Rural} & \textbf{Unspecified} & \textbf{Urban} \\
\midrule
Cosine similarity & 0.423$^{***}$ & 0.460$^{***}$ & 0.437$^{***}$ & 0.404$^{***}$ & 0.424$^{***}$ & 0.435$^{***}$ \\
Initial income & +1.8 (+19.0\%)$^{***}$ & +2.7 (+17.2\%)$^{***}$ & +3.1 (+17.8\%)$^{***}$ & +1.8 (+19.5\%)$^{***}$ & +2.8 (+19.7\%)$^{***}$ & +3.1 (+19.6\%)$^{***}$ \\
Initial savings & +6.7 (+72.3\%)$^{***}$ & +17.6 (+59.6\%)$^{*}$ & +23.0 (+77.4\%)$^{***}$ & +5.5 (+91.9\%)$^{***}$ & +13.4 (+78.7\%)$^{***}$ & +16.5 (+84.1\%)$^{***}$ \\
Initial total resources & +43.2 (+55.8\%)$^{***}$ & +108.5 (+50.7\%)$^{*}$ & +146.2 (+71.4\%)$^{***}$ & +36.3 (+64.7\%)$^{***}$ & +82.9 (+63.4\%)$^{***}$ & +114.4 (+77.6\%)$^{***}$ \\
Initial wealth & +34.7 (+59.2\%)$^{***}$ & +88.2 (+52.3\%)$^{*}$ & +120.2 (+76.1\%)$^{***}$ & +29.0 (+70.7\%)$^{***}$ & +66.8 (+66.9\%)$^{***}$ & +94.8 (+84.8\%)$^{***}$ \\
Post-advice income & +4.1 (+27.3\%)$^{***}$ & +5.7 (+23.2\%)$^{***}$ & +6.6 (+24.2\%)$^{***}$ & +3.8 (+26.8\%)$^{***}$ & +5.5 (+25.2\%)$^{***}$ & +5.9 (+23.5\%)$^{***}$ \\
Post-advice savings & +10.3 (+42.9\%)$^{***}$ & +21.9 (+38.5\%)$^{*}$ & +30.6 (+52.3\%)$^{***}$ & +9.1 (+51.0\%)$^{***}$ & +19.7 (+52.7\%)$^{***}$ & +22.0 (+50.5\%)$^{***}$ \\
Post-advice total resources & +62.3 (+43.9\%)$^{***}$ & +133.6 (+41.5\%)$^{**}$ & +174.1 (+53.2\%)$^{***}$ & +53.2 (+46.0\%)$^{***}$ & +105.7 (+47.1\%)$^{***}$ & +136.4 (+53.0\%)$^{***}$ \\
Post-advice wealth & +47.9 (+46.6\%)$^{***}$ & +106.0 (+44.0\%)$^{**}$ & +136.9 (+56.8\%)$^{***}$ & +40.2 (+48.2\%)$^{***}$ & +80.5 (+48.7\%)$^{***}$ & +108.5 (+57.5\%)$^{***}$ \\
\bottomrule
\end{tabular}
\end{adjustbox}
\vspace{0.35em}
{\tiny
\parbox{0.98\linewidth}{\textit{Note:} Financial outcomes are reported in units of 1,000 rupees. Cells report male-female differences; percentages are calculated as (male mean - female mean) / female mean within each employment group. * p < 0.05, ** p < 0.01, *** p < 0.001.}
}
\end{center}

\begin{center}
\tiny
\refstepcounter{table}
\textbf{Table \thetable: Average Male-Female Difference in Outcomes by Age}
\label{tab:Average-Age-Differences-in-Outcome}
\vspace{0.4em}
\begin{adjustbox}{max width=\linewidth, max totalheight=0.82\textheight, keepaspectratio}
\begin{tabular}{lrrrrrrrrr}
\toprule
 & \multicolumn{4}{c}{\textbf{Gender}} & \textbf{Gender x District} & \multicolumn{4}{c}{\textbf{Gender x Caste}} \\
\cmidrule(lr){2-5} \cmidrule(lr){6-6} \cmidrule(lr){7-10}
\textbf{Variable} & \textbf{18} & \textbf{25} & \textbf{30} & \textbf{70} & \textbf{25} & \textbf{18} & \textbf{25} & \textbf{30} & \textbf{70} \\
\midrule
Cosine similarity & 0.466$^{***}$ & 0.450$^{***}$ & 0.432$^{***}$ & 0.477$^{***}$ & 0.414$^{***}$ & 0.453$^{***}$ & 0.427$^{***}$ & 0.406$^{***}$ & 0.470$^{***}$ \\
Initial income & +0.3 (+6.0\%) & -0.1 (-0.7\%) & -0.5 (-2.8\%) & +5.2 (+35.1\%)$^{**}$ & +0.7 (+7.5\%)$^{***}$ & -0.0 (-1.0\%) & +0.6 (+4.7\%)$^{**}$ & +0.9 (+6.1\%)$^{***}$ & +3.4 (+27.0\%)$^{***}$ \\
Initial savings & -0.2 (-17.0\%) & -0.1 (-3.1\%) & -0.1 (-3.0\%) & +60.5 (+59.0\%)$^{*}$ & +0.0 (+0.2\%) & +0.0 (+0.9\%) & -0.1 (-3.2\%) & -0.1 (-4.6\%) & +53.0 (+92.5\%)$^{***}$ \\
Initial total resources & -0.6 (-5.2\%) & -1.1 (-3.6\%) & -2.3 (-5.7\%) & +362.3 (+54.2\%)$^{*}$ & +0.6 (+2.5\%)$^{***}$ & -1.0 (-9.2\%)$^{***}$ & -0.3 (-1.0\%) & -0.1 (-0.2\%) & +332.0 (+86.8\%)$^{***}$ \\
Initial wealth & -0.7 (-13.2\%) & -0.9 (-6.3\%) & -1.7 (-8.4\%) & +296.7 (+53.9\%) & -0.1 (-1.1\%)$^{**}$ & -1.0 (-17.7\%)$^{***}$ & -0.8 (-5.4\%)$^{**}$ & -0.8 (-4.2\%) & +275.6 (+88.2\%)$^{***}$ \\
Post-advice income & +1.5 (+16.8\%)$^{**}$ & +2.8 (+12.3\%)$^{**}$ & +2.7 (+10.1\%)$^{**}$ & +7.0 (+34.3\%)$^{***}$ & +3.2 (+19.4\%)$^{***}$ & +0.3 (+3.6\%) & +2.7 (+12.2\%)$^{***}$ & +4.3 (+17.7\%)$^{***}$ & +4.9 (+28.0\%)$^{***}$ \\
Post-advice savings & +0.2 (+4.1\%) & -0.8 (-7.4\%) & -1.5 (-10.7\%) & +70.4 (+43.4\%) & -0.0 (-0.6\%) & -0.1 (-1.7\%) & -0.3 (-3.3\%) & -1.4 (-10.7\%)$^{***}$ & +67.1 (+67.6\%)$^{***}$ \\
Post-advice total resources & +5.6 (+14.5\%)$^{*}$ & +3.3 (+3.6\%) & -0.9 (-0.7\%) & +399.9 (+48.7\%)$^{*}$ & +5.5 (+7.9\%)$^{***}$ & -0.8 (-2.0\%) & +2.3 (+2.6\%) & +3.5 (+3.1\%)$^{*}$ & +363.3 (+72.6\%)$^{***}$ \\
Post-advice wealth & +3.8 (+15.8\%)$^{*}$ & +1.2 (+2.1\%) & -2.1 (-2.8\%) & +322.4 (+50.5\%)$^{*}$ & +2.4 (+5.2\%)$^{***}$ & -1.0 (-4.1\%) & -0.1 (-0.1\%) & +0.6 (+0.8\%) & +291.3 (+76.0\%)$^{***}$ \\
\bottomrule
\end{tabular}
\end{adjustbox}
\vspace{0.35em}
{\tiny
\parbox{0.98\linewidth}{\textit{Note:}Financial outcomes are reported in units of 1,000 rupees. Cells report male-female differences; percentages are calculated as (male mean - female mean) / female mean within each employment group. * p < 0.05, ** p < 0.01, *** p < 0.001.}
}
\end{center}

% \subsubsection{Caste}

\begin{center}
\tiny
\refstepcounter{table}
\textbf{Table \thetable: Difference in Outcomes by Caste Across Gender, Employment, Education and Location}
\label{Difference-in-Outcomes-by-Caste-Across-Gender-Employment-Education-and-Location}
\vspace{0.4em}
\begin{adjustbox}{max width=\linewidth, max totalheight=0.82\textheight, keepaspectratio}
\begin{tabular}{lrrrrrrrrrrrrrr}
\toprule
\textbf{Variable} & \textbf{Man} & \textbf{Woman} & \textbf{Employed} & \textbf{Unemployed} & \textbf{Employment Unspecified} & \textbf{College Educated} & \textbf{High School Educated} & \textbf{Phd} & \textbf{Post-Graduate} & \textbf{Uneducated} & \textbf{Education Unspecified} & \textbf{Location} & \textbf{Rural} & \textbf{Urban} \\
\midrule
Initial wealth: General-SC & +96.8 (+94.5\%)$^{***}$ & +46.0 (+75.1\%)$^{***}$ & +113.2 (+73.7\%)$^{***}$ & +39.8 (+117.3\%)$^{***}$ & +61.3 (+105.3\%)$^{***}$ & +79.5 (+83.3\%)$^{***}$ & +43.2 (+91.8\%)$^{***}$ & +131.5 (+75.2\%)$^{***}$ & +120.3 (+90.0\%)$^{***}$ & +19.7 (+112.3\%)$^{***}$ & +34.5 (+151.4\%)$^{***}$ & +81.7 (+86.9\%)$^{***}$ & +27.9 (+65.8\%)$^{***}$ & +104.7 (+95.8\%)$^{***}$ \\
Initial income: General-SC & +2.7 (+19.0\%)$^{***}$ & +2.1 (+17.9\%)$^{***}$ & +7.3 (+18.7\%)$^{***}$ & -0.0 (-20.9\%) & +0.0 (+4.3\%) & +2.8 (+18.4\%)$^{***}$ & +1.8 (+18.2\%)$^{***}$ & +2.8 (+13.0\%)$^{***}$ & +3.1 (+17.2\%)$^{***}$ & +1.7 (+24.9\%)$^{***}$ & +2.4 (+31.9\%)$^{***}$ & +2.8 (+19.8\%)$^{***}$ & +1.4 (+15.1\%)$^{***}$ & +3.1 (+19.4\%)$^{***}$ \\
Initial savings: General-SC & +18.6 (+107.9\%)$^{***}$ & +8.8 (+90.1\%)$^{***}$ & +21.9 (+94.7\%)$^{***}$ & +8.0 (+124.8\%)$^{***}$ & +11.1 (+102.0\%)$^{***}$ & +15.4 (+92.5\%)$^{***}$ & +8.1 (+140.5\%)$^{***}$ & +27.3 (+90.0\%)$^{***}$ & +23.6 (+99.4\%)$^{***}$ & +2.7 (+152.0\%)$^{***}$ & +4.9 (+191.6\%)$^{***}$ & +16.2 (+101.8\%)$^{***}$ & +5.8 (+96.5\%)$^{***}$ & +19.1 (+102.8\%)$^{***}$ \\
Post-advice wealth: General-SC & +115.6 (+68.3\%)$^{***}$ & +60.9 (+52.7\%)$^{***}$ & +142.7 (+53.4\%)$^{***}$ & +48.8 (+73.8\%)$^{***}$ & +73.1 (+78.1\%)$^{***}$ & +96.9 (+59.9\%)$^{***}$ & +53.0 (+58.9\%)$^{***}$ & +164.4 (+58.4\%)$^{***}$ & +141.2 (+63.9\%)$^{***}$ & +28.4 (+63.1\%)$^{***}$ & +45.2 (+83.2\%)$^{***}$ & +100.1 (+63.0\%)$^{***}$ & +37.5 (+43.5\%)$^{***}$ & +127.0 (+69.8\%)$^{***}$ \\
Post-advice income: General-SC & +4.5 (+19.2\%)$^{***}$ & +3.4 (+18.2\%)$^{***}$ & +8.1 (+18.0\%)$^{***}$ & +1.8 (+18.9\%)$^{***}$ & +2.0 (+22.8\%)$^{***}$ & +4.8 (+20.5\%)$^{***}$ & +2.9 (+18.6\%)$^{***}$ & +5.4 (+15.1\%)$^{***}$ & +5.3 (+18.4\%)$^{***}$ & +2.3 (+20.3\%)$^{***}$ & +3.2 (+26.2\%)$^{***}$ & +4.8 (+21.2\%)$^{***}$ & +2.3 (+15.1\%)$^{***}$ & +4.9 (+18.8\%)$^{***}$ \\
Post-advice savings: General-SC & +28.3 (+76.6\%)$^{***}$ & +15.3 (+60.1\%)$^{***}$ & +33.6 (+64.8\%)$^{***}$ & +13.6 (+80.2\%)$^{***}$ & +18.2 (+73.4\%)$^{***}$ & +23.2 (+62.1\%)$^{***}$ & +14.8 (+87.5\%)$^{***}$ & +43.4 (+67.1\%)$^{***}$ & +34.6 (+68.9\%)$^{***}$ & +5.6 (+68.0\%)$^{***}$ & +9.3 (+94.9\%)$^{***}$ & +26.0 (+73.8\%)$^{***}$ & +11.3 (+65.1\%)$^{***}$ & +28.2 (+68.6\%)$^{***}$ \\
Initial wealth: General-ST & +97.4 (+95.6\%)$^{***}$ & +44.4 (+70.6\%)$^{***}$ & +111.6 (+71.9\%)$^{***}$ & +40.2 (+120.2\%)$^{***}$ & +60.9 (+104.0\%)$^{***}$ & +69.7 (+66.3\%)$^{***}$ & +44.3 (+96.3\%)$^{***}$ & +132.8 (+76.5\%)$^{***}$ & +123.6 (+94.8\%)$^{***}$ & +18.4 (+98.2\%)$^{***}$ & +36.7 (+177.8\%)$^{***}$ & +79.8 (+83.2\%)$^{***}$ & +30.2 (+75.4\%)$^{***}$ & +102.7 (+92.2\%)$^{***}$ \\
Initial income: General-ST & +3.0 (+21.7\%)$^{***}$ & +2.0 (+17.1\%)$^{***}$ & +7.6 (+19.7\%)$^{***}$ & +0.0 (+12.8\%) & +0.0 (+5.3\%) & +2.9 (+19.0\%)$^{***}$ & +1.9 (+19.2\%)$^{***}$ & +3.4 (+16.3\%)$^{***}$ & +3.0 (+16.8\%)$^{***}$ & +1.7 (+24.3\%)$^{***}$ & +2.5 (+32.5\%)$^{***}$ & +3.0 (+21.4\%)$^{***}$ & +1.9 (+20.9\%)$^{***}$ & +2.8 (+17.3\%)$^{***}$ \\
Initial savings: General-ST & +18.7 (+109.1\%)$^{***}$ & +8.0 (+76.9\%)$^{***}$ & +21.6 (+91.8\%)$^{***}$ & +7.9 (+123.1\%)$^{***}$ & +10.6 (+92.6\%)$^{***}$ & +13.5 (+72.6\%)$^{***}$ & +8.2 (+144.5\%)$^{***}$ & +26.4 (+84.4\%)$^{***}$ & +24.3 (+105.6\%)$^{***}$ & +2.6 (+129.6\%)$^{***}$ & +5.2 (+237.0\%)$^{***}$ & +15.7 (+95.5\%)$^{***}$ & +5.9 (+100.4\%)$^{***}$ & +18.6 (+97.1\%)$^{***}$ \\
Post-advice wealth: General-ST & +116.6 (+69.4\%)$^{***}$ & +58.0 (+49.0\%)$^{***}$ & +139.1 (+51.3\%)$^{***}$ & +50.7 (+78.8\%)$^{***}$ & +72.2 (+76.6\%)$^{***}$ & +86.3 (+50.0\%)$^{***}$ & +52.7 (+58.2\%)$^{***}$ & +165.2 (+58.8\%)$^{***}$ & +143.6 (+65.7\%)$^{***}$ & +27.7 (+60.4\%)$^{***}$ & +48.5 (+95.2\%)$^{***}$ & +99.7 (+62.5\%)$^{***}$ & +39.6 (+47.1\%)$^{***}$ & +122.7 (+65.9\%)$^{***}$ \\
Post-advice income: General-ST & +5.1 (+22.0\%)$^{***}$ & +3.4 (+18.1\%)$^{***}$ & +8.4 (+18.9\%)$^{***}$ & +2.2 (+23.0\%)$^{***}$ & +2.1 (+24.1\%)$^{***}$ & +4.9 (+21.1\%)$^{***}$ & +3.1 (+20.1\%)$^{***}$ & +6.0 (+17.2\%)$^{***}$ & +5.3 (+18.3\%)$^{***}$ & +2.5 (+22.3\%)$^{***}$ & +3.6 (+31.0\%)$^{***}$ & +5.1 (+23.2\%)$^{***}$ & +3.0 (+20.3\%)$^{***}$ & +4.6 (+17.7\%)$^{***}$ \\
Post-advice savings: General-ST & +28.5 (+77.3\%)$^{***}$ & +14.5 (+55.3\%)$^{***}$ & +33.1 (+63.1\%)$^{***}$ & +13.6 (+79.4\%)$^{***}$ & +17.8 (+70.9\%)$^{***}$ & +20.6 (+51.4\%)$^{***}$ & +14.5 (+84.0\%)$^{***}$ & +42.8 (+65.6\%)$^{***}$ & +35.3 (+71.2\%)$^{***}$ & +5.4 (+64.7\%)$^{***}$ & +10.4 (+118.1\%)$^{***}$ & +25.3 (+70.2\%)$^{***}$ & +10.9 (+61.7\%)$^{***}$ & +28.3 (+69.1\%)$^{***}$ \\
Initial wealth: General-OBC & +67.8 (+51.6\%)$^{***}$ & +24.8 (+30.1\%)$^{***}$ & +79.3 (+42.3\%)$^{***}$ & +27.7 (+60.4\%)$^{***}$ & +32.0 (+36.5\%)$^{***}$ & +39.6 (+29.3\%)$^{***}$ & +24.9 (+38.1\%)$^{***}$ & +104.3 (+51.6\%)$^{***}$ & +74.2 (+41.2\%)$^{**}$ & +12.0 (+47.6\%)$^{***}$ & +23.0 (+66.9\%)$^{***}$ & +48.7 (+38.4\%)$^{***}$ & +16.6 (+30.9\%)$^{***}$ & +73.7 (+52.5\%)$^{***}$ \\
Initial income: General-OBC & +1.7 (+11.1\%)$^{***}$ & +1.0 (+8.0\%)$^{***}$ & +4.1 (+9.8\%)$^{***}$ & -0.0 (-26.4\%) & +0.0 (+23.7\%) & +1.4 (+8.5\%)$^{***}$ & +0.9 (+8.8\%)$^{***}$ & +2.1 (+9.6\%)$^{***}$ & +1.8 (+9.6\%)$^{***}$ & +0.9 (+11.6\%)$^{***}$ & +1.1 (+11.9\%)$^{**}$ & +1.4 (+9.1\%)$^{***}$ & +0.9 (+8.9\%)$^{***}$ & +1.8 (+10.7\%)$^{***}$ \\
Initial savings: General-OBC & +13.8 (+62.7\%)$^{***}$ & +4.8 (+34.8\%)$^{***}$ & +16.6 (+58.6\%)$^{***}$ & +5.5 (+62.0\%)$^{***}$ & +5.7 (+35.1\%)$^{***}$ & +8.4 (+35.5\%)$^{***}$ & +5.2 (+60.6\%)$^{***}$ & +21.7 (+60.4\%)$^{***}$ & +15.4 (+48.1\%)$^{**}$ & +1.7 (+58.2\%)$^{**}$ & +3.3 (+81.4\%)$^{***}$ & +10.5 (+48.8\%)$^{***}$ & +3.8 (+47.4\%)$^{***}$ & +13.6 (+56.4\%)$^{***}$ \\
Post-advice wealth: General-OBC & +79.6 (+38.8\%)$^{***}$ & +31.6 (+21.8\%)$^{***}$ & +97.5 (+31.2\%)$^{***}$ & +31.6 (+37.9\%)$^{***}$ & +37.6 (+29.2\%)$^{***}$ & +49.1 (+23.4\%)$^{***}$ & +28.3 (+24.6\%)$^{***}$ & +126.0 (+39.4\%)$^{***}$ & +83.4 (+29.9\%)$^{**}$ & +17.0 (+30.1\%)$^{***}$ & +29.6 (+42.3\%)$^{***}$ & +59.8 (+30.0\%)$^{***}$ & +21.9 (+21.5\%)$^{***}$ & +85.1 (+38.0\%)$^{***}$ \\
Post-advice income: General-OBC & +2.8 (+11.0\%)$^{***}$ & +1.6 (+7.6\%)$^{***}$ & +4.7 (+9.7\%)$^{***}$ & +0.9 (+8.0\%)$^{***}$ & +1.0 (+9.9\%)$^{***}$ & +2.5 (+9.5\%)$^{***}$ & +1.6 (+9.4\%)$^{***}$ & +3.4 (+8.9\%)$^{***}$ & +3.3 (+10.6\%)$^{***}$ & +1.1 (+9.0\%)$^{***}$ & +1.3 (+8.9\%)$^{***}$ & +2.4 (+9.7\%)$^{***}$ & +1.3 (+8.3\%)$^{***}$ & +2.8 (+9.9\%)$^{***}$ \\
Post-advice savings: General-OBC & +20.9 (+46.9\%)$^{***}$ & +8.9 (+27.8\%)$^{***}$ & +26.6 (+45.1\%)$^{***}$ & +8.7 (+39.5\%)$^{***}$ & +9.3 (+27.8\%)$^{***}$ & +12.8 (+26.8\%)$^{***}$ & +9.2 (+41.2\%)$^{***}$ & +34.3 (+46.6\%)$^{***}$ & +23.0 (+37.2\%)$^{***}$ & +3.6 (+35.9\%)$^{***}$ & +6.2 (+47.6\%)$^{***}$ & +17.4 (+39.6\%)$^{***}$ & +7.7 (+37.0\%)$^{***}$ & +19.5 (+39.2\%)$^{***}$ \\
\bottomrule
\end{tabular}

\end{adjustbox}
\vspace{0.35em}
{\tiny
\parbox{0.98\linewidth}{\textit{Note:} Financial outcomes are reported in units of 1,000 rupees. Cells report male-female differences; percentages are calculated as (male mean - female mean) / female mean within each employment group. * p < 0.05, ** p < 0.01, *** p < 0.001.}
}
\end{center}

\begin{center}
\tiny
\refstepcounter{table}
\textbf{Table \thetable: Gender Coefficients from Financial Outcome Regressions}
\label{tab:Gender-Coefficients-from-Financial-Outcome-Regressions}
\vspace{0.4em}
\begin{adjustbox}{max width=\linewidth, max totalheight=0.82\textheight, keepaspectratio}
\begin{tabular}{lrrrrrr}
\toprule
 & \multicolumn{2}{c}{\textbf{Gender}} & \multicolumn{2}{c}{\textbf{Gender x District}} & \multicolumn{2}{c}{\textbf{Gender x Caste}} \\
\cmidrule(lr){2-3} \cmidrule(lr){4-5} \cmidrule(lr){6-7}
\textbf{Variable} & \textbf{Coefficient} & \textbf{N} & \textbf{Coefficient} & \textbf{N} & \textbf{Coefficient} & \textbf{N} \\
\midrule
Income gain & 2.7***
(0.3) & 520 & 1.8
(nan) & 12432 & 2.5***
(0.1) & 7020 \\
Initial income & 4.1***
(0.7) & 520 & 0.8
(5,914.8) & 12432 & 2.5***
(0.1) & 7020 \\
Initial savings & 20.6***
(5.7) & 520 & 0.0
(1,208.1) & 12432 & 11.8***
(0.9) & 7020 \\
Initial total resources & 137.0***
(34.2) & 520 & 1.0
(12,751.1) & 12432 & 77.9***
(5.6) & 7020 \\
Initial wealth & 112.4***
(28.3) & 520 & 0.2
(9,698.2) & 12432 & 63.5***
(4.7) & 7020 \\
Post-advice income & 6.7***
(0.8) & 520 & 2.6
(nan) & 12432 & 5.1***
(0.1) & 7020 \\
Post-advice savings & 29.1***
(6.8) & 520 & 0.2
(3,012.1) & 12432 & 17.0***
(1.2) & 7020 \\
Post-advice total resources & 169.0***
(37.5) & 520 & 5.0
(32,333.8) & 12432 & 98.4***
(6.3) & 7020 \\
Post-advice wealth & 133.2***
(30.5) & 520 & 2.2
(11,817.5) & 12432 & 76.4***
(5.1) & 7020 \\
Savings gain & 8.5***
(1.9) & 520 & 0.2
(904.9) & 12432 & 5.2***
(0.4) & 7020 \\
Total gain & 32.0***
(5.7) & 520 & 4.0
(nan) & 12432 & 20.6***
(1.1) & 7020 \\
Wealth gain & 20.8***
(4.1) & 520 & 2.0
(nan) & 12432 & 12.9***
(0.8) & 7020 \\
\bottomrule
\end{tabular}
\end{adjustbox}
\vspace{0.35em}
{\tiny
\parbox{0.98\linewidth}{\textit{Note:} Values in brackets are standard errors.* p < 0.05, ** p < 0.01, *** p < 0.001.}
}
\end{center}

\begin{figure}[H]
\centering

\end{figure}
%%%%%%%%%%%%%%%%%%%%%%%%%%%%%%%%%%%%%%%%%%%%%%%%%%%%%%%%%%%%

% \appendix

% \section{Technical appendices and supplementary material}
% Technical appendices with additional results, figures, graphs, and proofs may be submitted with the paper submission before the full submission deadline (see above). You can upload a ZIP file for videos or code, but do not upload a separate PDF file for the appendix. There is no page limit for the technical appendices. 

% Note: Think of the appendix as ``optional reading'' for reviewers. The paper must be able to stand alone without the appendix; for example, adding critical experiments that support the main claims to an appendix is inappropriate. 

%%%%%%%%%%%%%%%%%%%%%%%%%%%%%%%%%%%%%%%%%%%%%%%%%%%%%%%%%%%%

% \newpage
% \input{checklist.tex}

\end{document}